\documentclass[lettersize,journal]{IEEEtran}
\usepackage[utf8]{inputenc}
\usepackage[nolist]{acronym}
\usepackage{amsmath}

\usepackage{mathtools}
\usepackage[linesnumbered,ruled,vlined]{algorithm2e}
\usepackage{graphics}
\usepackage{breqn}
\usepackage{amsmath, amssymb}
\usepackage{algorithmic}
\usepackage{cite}
\usepackage{graphicx}
\graphicspath{{Figures/}}
\usepackage{xcolor} 

\usepackage{subcaption}

\begin{document}
\title{Scalable machine learning-based approaches for energy saving in densely deployed Open RAN}

\author{\IEEEauthorblockN{Xuanyu Liang\IEEEauthorrefmark{1}, Ahmed Al-Tahmeesschi\IEEEauthorrefmark{1}, Swarna Chetty\IEEEauthorrefmark{1}, Cicek Cavdar\IEEEauthorrefmark{2}, Berk Canberk \IEEEauthorrefmark{3}
and Hamed Ahmadi\IEEEauthorrefmark{1}}
\\
\IEEEauthorrefmark{1}School of Physics Engineering and Technology, University of York, UK\\
\IEEEauthorrefmark{2}School of EECS at KTH Royal Institute of Technology, Sweden.\\
\IEEEauthorrefmark{3}School of Computing, Engineering and The Built Environment, Edinburgh Napier University, UK\\
}

\maketitle

\begin{abstract}

Densely deployed base stations are responsible for the majority of the energy consumed in Radio access network (RAN). While these deployments are crucial to deliver the required data rate in busy hours of the day, the network can save energy by switching some of them to sleep mode and maintain the coverage and quality of service with the other ones. Benefiting from the flexibility provided by the Open RAN in embedding machine learning (ML) in network operations, in this work we propose Deep Reinforcement Learning (DRL)-based energy saving solutions. Firstly we propose 3 different DRL-based methods in the form of xApps which control the Active/Sleep mode of up to 6 radio units (RUs) from Near Real time RAN Intelligent Controller (RIC). We also propose a further scalable federated DRL-based solution with an aggregator as an rApp in None Real time RIC and local agents as xApps. Our simulation results present the convergence of the proposed methods. We also compare the performance of our federated DRL across three layouts spanning 6--24 RUs and 500--1000\,m regions, including a composite multi-region scenario. The results show that our proposed federated TD3 algorithm achieves up to 43.75\% faster convergence, more than 50\% network energy saving and 37. 4\% lower training energy versus centralized baselines, while maintaining the quality of service and improving the robustness of the policy.

\end{abstract}

\section{Introduction}
To accommodate rapidly increasing mobile traffic, \acp{BS} have been widely deployed to satisfy user data rate requirements. Although dense deployments significantly enhance capacity and coverage, they also lead to substantial energy consumption. Studies indicate that \acp{BS} account for a major portion of total network energy usage: roughly a decade ago they drew about 57\% of the total, whereas recent figures have risen to 73\%–77\%~\cite{thompson2022editorial}. This firmly establishes \acp{BS} as the dominant energy consumers in mobile networks. A typical 5G macro \ac{BS} consumes approximately $3.3$–$9$~kW, with higher values observed for massive MIMO and mmWave deployments. The resulting energy demand increases both the operational expenditure of network providers and the associated greenhouse gas emissions~\cite{ahmadi2025towards}. Consequently, improving energy efficiency in BS operations is increasingly crucial. However, traditional \ac{BS} designs feature tightly integrated hardware with limited flexibility, constraining selective activation or power adaptation of individual components~\cite{polese2023understanding} and leading to persistently inefficient energy usage.

In this context, \ac{O-RAN} introduces an open, disaggregated, and intelligence-ready architecture that better supports energy-aware control through standardized interfaces~\cite{O-RanArchitecture}. It decomposes the monolithic \ac{BS} into the \ac{CU}, \ac{DU}, and \ac{RU}; the \ac{CU}/\ac{DU} together implement \ac{BBU} functions and are typically cloud-hosted, while the \ac{RU} remains a physical entity (akin to a streamlined \ac{RRH}) responsible for lower-PHY processing, RF chains, and power amplifiers—components that constitute a major share of O-RAN energy consumption~\cite{abubakar2023energy}. As \acp{RU} are dimensioned for peak demand, placing a larger fraction into sleep mode during off-peak periods can yield substantial savings provided \ac{QoS} constraints are respected~\cite{lahdekorpi2017energy}. O-RAN further introduces the \ac{RIC} as a programmable control platform with two layers: the \ac{Near-RT RIC} (control loop $\sim$10--1000~ms) hosting third-party xApps for near-real-time radio control (e.g., traffic steering \cite{dryjanski2021toward}, network slicing \cite{bonati2021intelligence} and latency sensitive control \cite{zhou2025digital}) and the \ac{Non-RT RIC} (timescales $\geq$1~s) orchestrating policies and analytics through rApps~\cite{agarwal2025open}. 
The Non-RT RIC provides high-level guidance and long-term optimization goals to the Near-RT RIC via the standardized A1 interface \cite{O-RanA1Interface}. 
In turn, the Near-RT RIC executes time-sensitive control by running xApps that translate these A1 policies into actionable decisions and interact with O-DU/O-RU nodes through the E2 interface\cite{O-RanE2Interface}. 
This hierarchical workflow enables rApps to supply long-term network-wide intelligence (e.g., predicted traffic trends or RU utilization expectations), while xApps react to instantaneous conditions to enforce fine-grained radio control. The whole workflow is illustrated at Fig. \ref{fig:systemmodel1}.
Building on this architecture, AI-driven solutions are increasingly deployed in O-RAN to support intelligent and adaptive control at both rApps and xApps \cite{zhou2025integrating}. Several prior studies and O-RAN specification documents have analyzed signaling latency, control overhead, and RIC interaction models under similar functional splits. Our framework adheres to these established timelines and control abstractions. In our work, we develop decentralized \ac{ML}-based mechanisms in the form of an xApp that leverages rApp-provided policy constraints to intelligently schedule \acp{RU} into sleep mode while maintaining user \ac{QoS}.

\subsection{Related Work}
Early efforts to reduce energy consumption in cellular networks focused on \ac{BS} sleep-mode control. When \acp{BS} are densely deployed for peak traffic, keeping them active during off-peak periods leads to substantial energy waste. Turning off lightly loaded sites---a strategy commonly referred to as \emph{sleep mode}---has therefore been widely studied \cite{wu2015energy,liu2015small,oh2016unified,peng2014stochastic}. For instance, \cite{liu2015small} introduced a random sleep policy for small-cell \acp{BS}, while \cite{oh2016unified} proposed a sequential deactivation algorithm that preserves user rates. In \cite{peng2014stochastic}, several schemes were analyzed for macro-\ac{BS} deactivation in heterogeneous networks, maintaining coverage via power readjustment and complementary micro layers. Although such optimization-based approaches can be effective, their scalability degrades with growing network size and multi-objective constraints, leading to high computational complexity and long solution times.

The sleep/active mode decision problem is NP-hard due to the combinatorial on/off choices and temporal coupling, which makes exact optimization intractable at scale. To address this challenge while satisfying \ac{QoS} constraints, \ac{ML}-based strategies have been widely studied, particularly in \ac{UDN} scenarios where spatial redundancy enables flexible sleeping. Paper~\cite{jang2020base} employs an \ac{LSTM}-based predictor to capture temporal patterns in traffic and channel variations, enabling proactive on/off switching. In~\cite{ju2022energy}, a \ac{DQN} framework is enhanced with an action-selection network that filters out invalid actions and mitigates ineffective exploration. The work in~\cite{ye2019drag} introduces a modular \ac{DRL} pipeline that integrates a \ac{DDPG} policy with a supervised cost estimator and traffic predictor to account for switching costs and \ac{QoS} degradation.  
Spatio–temporal correlations are further leveraged in~\cite{wu2021deep}, where convolutional-\ac{LSTM} forecasting is coupled with an actor–critic controller to improve robustness under fluctuating traffic. A different direction is explored in~\cite{gan2025joint}, which formulates joint sleep control and renewable-energy sharing as a single \ac{MDP}, solved through a multi-discrete \ac{PPO} that factorizes the exponential action space into tractable subspaces. Paper~\cite{sun2024deep} presents a \ac{PPO}-based approach that simultaneously considers sleep scheduling, cell zooming, user association, and \ac{RIS} configuration, thus addressing multi-cell coordination under mixed discrete/continuous actions.  
Risk-aware mechanisms have also been investigated. In~\cite{masoudi2022digital}, a digital twin is integrated with a \ac{DQN} controller to assess delay risk before executing sleep decisions, triggering re-training or feature suppression under anomalous traffic conditions. From a multi-agent perspective,~\cite{zhen2024energy} treats each \ac{BS} as an agent and applies a multi-agent \ac{PPO} with a state-similarity heuristic, jointly optimizing BS sleeping and MIMO antenna operations while reducing oscillatory behavior.
Recent works have explored heuristic and learning-based \ac{RC} switching strategies within the \ac{O-RAN} framework. For example, \cite{liang2024energy} proposed heuristic xApps for \ac{RC} ON–OFF control under static user deployment, where \acp{UE} were assumed to be non-mobile and decisions were made based on instantaneous load and RSS thresholds. While such approaches demonstrate energy savings in snapshot-like scenarios, the decision rules are manually designed and rely on fixed thresholds.
Subsequently, \cite{wang2024energy} introduced a DQN-based xApp under a similar static network setting and showed that learning-based policies outperform the heuristic model, even in stationary environments. This indicates that the \ac{RC} activation problem is inherently combinatorial and difficult to fully capture through fixed rule-based logic.
Both lines of work primarily consider static or quasi-static user distributions. In practical deployments, user mobility and traffic fluctuations introduce temporal dependencies, transforming the RC switching problem from a one-shot optimization into a sequential control problem with long-term consequences.

However, centralized \ac{ML} approaches typically collect training data at a single server, which raises scalability concerns in large, distributed systems: as \acp{RU}/\acp{BS} and \acp{UE} increase, the communication burden grows rapidly and data heterogeneity across locations hinders convergence and generalization \cite{zhou2023toward}. \ac{FL} addresses these issues by training locally and aggregating model updates instead of raw data, reducing backhaul load and improving locality adaptation \cite{zhao2025multi}. Nevertheless, \ac{FRL} is sensitive to client heterogeneity; inconsistent local policies can lead to conflicting updates and a less effective global policy \cite{jin2022federated}.

Applying federated \ac{DRL} to sleep control has shown promise. For example, \cite{pan2024dynamic} lets each \ac{SBS} train a local \ac{DDQN} and periodically aggregates at a macro \ac{BS}. It treats each \ac{SBS} as an isolated agent, resembling a loosely coupled multi-agent system; and it aggregates at the macro-site level, which is agnostic to the functional split and control timescales defined in O-RAN. In contrast to prior cellular-centric FL/FRL designs, O-RAN introduces a native control hierarchy (Non-RT RIC for long-timescale policy/orchestration and Near-RT RIC for near-real-time control). Moreover, \acp{RU} are major energy consumers in O-RAN due to RF and PA hardware. Our work aligns the learning workflow with this hierarchy and shifts the agent granularity from per-\ac{SBS} to per-region (i.e., an area comprising multiple \acp{RU} and \acp{UE}). This broader spatial/temporal context improves policy generalization across heterogeneous regions while remaining faithful to O-RAN interfaces and roles. As a result, the proposed federated deep reinforcement learning framework aggregates richer, more representative local policies and scales better than centralized training in large, distributed O-RAN deployments.

\subsection{Contributions}
In this paper, we investigate the energy-efficient operation in the O-RAN architecture, which, due to provisioning for peak demand, often keeps \acp{RU} active even during low traffic periods, leading to unnecessary energy consumption. We formulate the joint RU sleep control problem as an NP-hard network-wide energy minimization task under \ac{QoS} constraints, and transform it into a \ac{MDP} to enable sequential decision-making via \ac{DRL}. To address the challenge of large action spaces, we develop three centralized DRL schemes: (i) a \ac{DQNMA} baseline, which enumerates all joint RU configurations; (ii) a \ac{DQNSA} variant, which sequentially switches the state of only one RU per step; and (iii) a \ac{TD3} controller, which outputs continuous control values, thereby avoiding exponential action growth. Although these centralized approaches are effective in small or medium scale deployments, their scalability is limited in large and geographically distributed networks. To overcome this, The proposed Fed-DRL architecture provides scalable learning across large O-RAN deployments by enabling each area to train its policy locally while only exchanging neural-network parameters with a global aggregator. This avoids raw data sharing, eliminates the computational bottleneck of centralized training. During execution, each xApp independently makes RU activation decisions based on local observations, further improving scalability.
The main contributions of this work are summarized as follows:
\begin{itemize}
    \item We propose an O-RAN compliant energy control framework in which \acp{RU} dynamically switch between active and sleep modes based on traffic and mobility patterns. The design leverages the O-RAN functional split and standardized control interfaces for practical deployment.
    \item We formulate the activation of \ac{RU} as \ac{MDP} considering traffic load, user mobility, and spatial distribution. The reward function incorporates empirically validated energy and \ac{QoS} tradeoffs to ensure stable and meaningful learning behavior. This formulation is compatible with the O-RAN learning framework aligned with multiple \ac{DRL} paradigms, supporting both centralized and distributed solutions.
    \item We propose a federated DRL architecture tailored for O-RAN, where local agents embedded in the Near-RT RIC interact with regional \acp{RU} and \acp{UE}, and their models are periodically aggregated at the Non-RT RIC. This architecture is consistent with the disaggregated control roles of O-RAN, enabling real-time local responsiveness while ensuring global policy consistency and coordination across regions. By explicitly embedding FL into the O-RAN control framework, our design provides a scalable and communication-efficient solution for large and heterogeneous network deployments.
    \item We conduct extensive simulations across multiple network layouts and heterogeneous regions. The results show that the proposed Fed-DRL achieves significant network energy savings and scales effectively to large deployments. Moreover, the federated-learning–based solution also reduces the computational and communication overhead associated with centralized training, thereby lowering the overall training energy consumption.
\end{itemize}



\section{System Model}





\begin{figure*}[!htbp]
	\centering
	\includegraphics[clip, trim=0.0cm 0.cm 0.0cm 0cm, width=.8\textwidth]{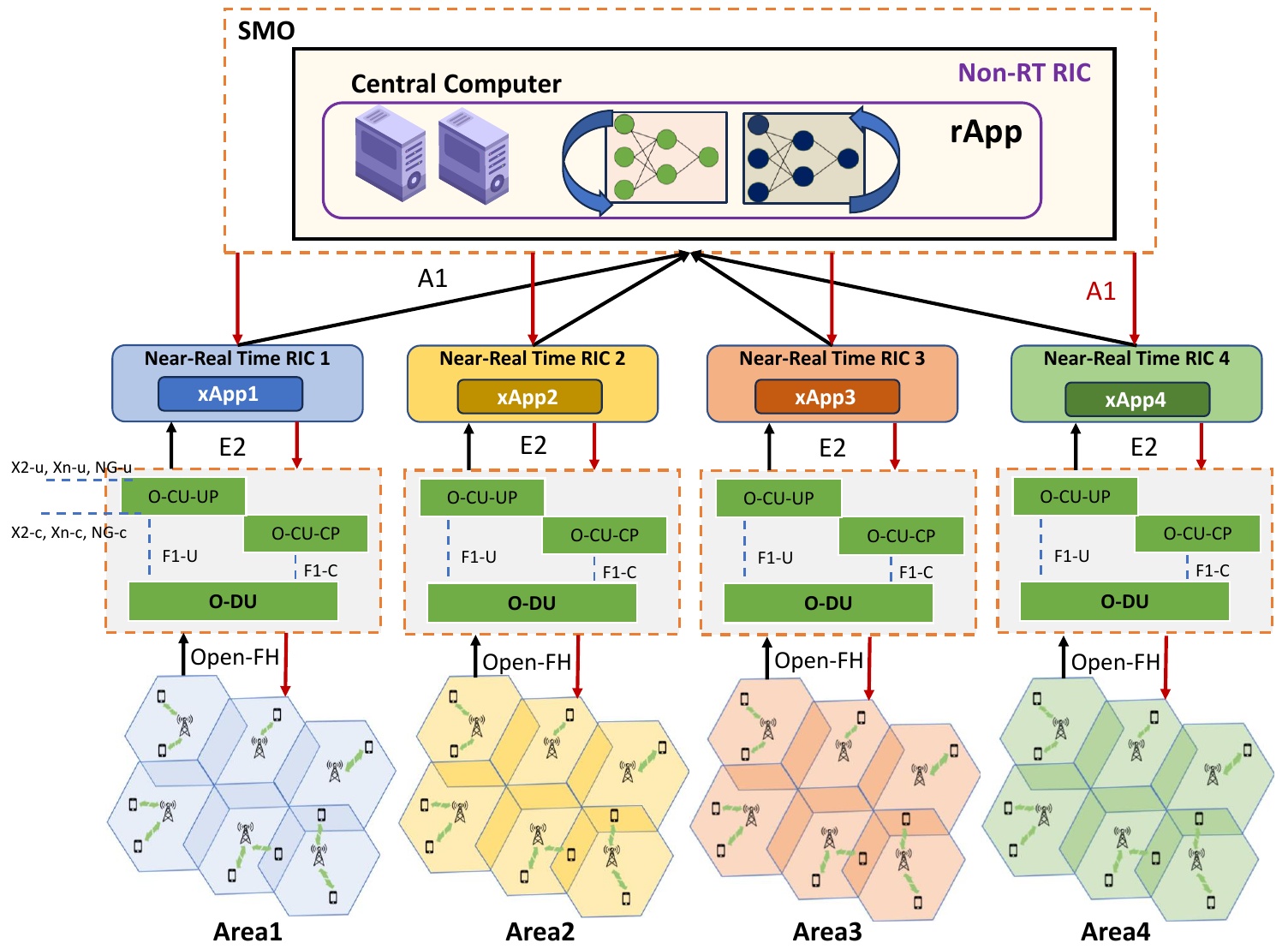}
	\caption{Illustration of the O-RAN architecture incorporating both O-RAN-defined and 3GPP-standard interfaces. Solid lines represent O-RAN interfaces (e.g., E2, A1, O1, Open FH), while dashed lines indicate 3GPP interfaces.The system spans four geographical areas, each with its own O-DU and O-CU components. Near-RT RICs operate on a per-area basis, deploying multiple xApps. A centralized Non-RT RIC performs global policy aggregation and training coordination using A1 interface.}
	\label{fig:systemmodel1}
\end{figure*}

In this section, we describe the system model of the O-RAN-based wireless network under consideration. The network consists of $M$ distributed \acp{RU}, each equipped with a single antenna, serving $K$ single-antenna \acp{UE} in the downlink. The sets of RUs and UEs are denoted by $\mathcal{M} = \{1, 2, \ldots, M\}$ and $\mathcal{K} = \{1, 2, \ldots, K\}$, respectively.
We assume a multi-area deployment, where RUs are geographically distributed and managed by corresponding near-real-time RIC instances as illustrated on Fig. \ref{fig:systemmodel1}. The system operates in discrete time slots and the UEs are mobile with a constant speed $v$, following a random direction mobility model. Each UE is associated with one RU based on signal strength and proximity. To model energy control at the RU level, we define a binary variable $\alpha_m \in \{0,1\}$ for each RU $m \in \mathcal{M}$. Specifically, $\alpha_m = 1$ indicates that the RU $m$ is in active mode and capable of transmitting data, while $\alpha_m = 0$ represents that the RU is in sleep mode and does not consume transmission power during the interval.

In our system, each \ac{RU} $m \in \mathcal{M}$ is equipped with a total of $Q_m$ \acp{PRB}, which can be allocated to its associated \acp{UE}. Let $U_m^t$ denote the number of \acp{UE} associated with \ac{RU} $m$ at time slot $t$. We define $n_{m,k}^t$ as the number of \acp{PRB} allocated by RU $m$ to the $k$-th \ac{UE} at time $t$, where $k \in \mathcal{K}$.
Based on this allocation, the total number of \acp{PRB} utilized by RU $m$ at time $t$, denoted as $N_m^t$, can be expressed as:
\begin{equation}
N_m^t = \sum_{k = 1}^{U_m^t} n_{m,k}^t.
\end{equation}
The \ac{PRB} allocation $n_{m,k}^t$ for each UE is determined according to its individual data rate requirement at time $t$, ensuring that the allocated resources are sufficient to meet the minimum \ac{QoS} demands.
Based on the PRB allocation, we define the load of RU $m$ at time slot $t$ as the ratio of occupied PRBs to the total available PRBs:

\begin{equation} \label{eq:ru_load}
l_{m}^{t} = \frac{N_{m}^{t}}{Q_m}.
\end{equation}
To model the wireless channel, we adopt the \ac{UMi} fading model. Let $h_{m,k}^t$ denote the small-scale fading channel gain between RU $m$ and UE $k$ at time slot $t$. We consider a power allocation scheme where the total transmission power of each RU, denoted as $P_{\mathrm{TX}}$, is evenly distributed across all its available \acp{PRB}. Consequently, the transmission power allocated to a \ac{UE} is proportional to the number of \acp{PRB} it receives.
Under this model, the received \ac{SNR} at \ac{UE} $k$ served by RU $m$ at time $t$ is given by:

\begin{equation}
SNR_{m,k}^{t} = \frac{P_{\mathrm{TX}} \cdot h_{m,k}^{t} \cdot \left(\frac{n_{m,k}^{t}}{Q_m}\right)}{n_{m,k}^{t} \cdot N_0},
\end{equation}
where $N_0$ represents the noise power spectral density of the \ac{AWGN}, and the term $\frac{n_{m,k}^{t}}{Q_m}$ reflects the fraction of total RU power allocated to \ac{UE} $k$ based on its \ac{PRB} share.
The achievable downlink data rate for \ac{UE} $k$ at time $t$ is then expressed by the Shannon capacity formula as:

\begin{equation}
R_{k}^{t} = n_{m,k}^{t} \cdot B \cdot \log_2\left(1 + SNR_{m,k}^{t}\right),
\end{equation}
where $B$ denotes the bandwidth of a single PRB.

\begin{table}[h]
\centering
\caption{Summary of Symbols}
\begin{tabular}{|l|l|}
\hline
\textbf{Notations}       & \textbf{Value} \\ \hline
$\alpha_m$               & Mode variable of \ac{RU} $m$ \\ \hline
$n_m,k$              & Number of allocated \acp{PRB} from \ac{RU}$m$ to \ac{UE} $k$ \\\hline
$N_m$                    & Number of \acp{PRB} allocated in \ac{RU} $m$              \\ \hline
$Q_m$                    & Total number of \acp{PRB} in \ac{RU} $m$\\ \hline
$U_m^t$                  & Total number of \acp{UE} associate with \ac{RU} $m$ at time slot $t$\\ \hline
$l_m$                    & Load of \ac{RU} $m$               \\ \hline
$h_{m,k}$                & UMi fading channel model from the \ac{RU} $m$ to the \ac{UE} $k$\\ \hline
$SNR_{m,k}$              & \ac{SNR} of the \ac{UE} $m$ in \ac{RU} $k$             \\ \hline
$R_{d,k,min}$              & Data rate requirement of \ac{UE} $k$               \\ \hline
$R_{d,k}$                  & Current data rate requirement of \ac{UE} $k$ \\ \hline
$\eta$                   & Power amplifier efficiency               \\ \hline
$P_{TX}$                 & Maximum transmission power of RU               \\ \hline
$V_m^{trans}$            & Mode transition power of \ac{RU}  $m$               \\ \hline
$P_{m}^{t}$             & Energy consumption of RU $m$ at time slot $t$\\ \hline
$P_m^{active}$          & Fixed power consumption of RU $m$ in active mode\\ \hline
$P_m^{sleep}$           & Fixed power consumption of RU $m$ in sleep mode\\ \hline
$P_m^{data}$           & Load dependent power consumption of RU $m$\\ \hline
$P_{\text{tot}}^t$           & Total network energy consumption at time slot $t$\\ \hline
\end{tabular}
\end{table}

\subsection{Power Consumption Model and Problem Formulation}
The power consumption of each RU is composed of three components~\cite{ju2022energy}. The first is fixed power consumption, denoted by $P_{m}^{\mathrm{Fix},t}$, which represents the energy used by signal processing, cooling systems, and power supply units. This component depends solely on the operational mode of the RU. 
The second component is the load-dependent power consumption, denoted by $P_{m}^{\mathrm{data},t}$, primarily attributed to the \ac{PA}. In this work, the load is quantified based on the \ac{PRB} utilization, as defined in~\eqref{eq:ru_load}. 
The third component is the transition power $P_{m}^{\mathrm{trans},t}$, which is incurred only when the \ac{RU} switches from sleep mode to active mode. This represents the additional energy required to re-activate hardware components such as the signal processing unit and power amplifier.
By summing these three components, the total power consumption of \ac{RU} $m$ at time slot $t$ is given by:
\begin{equation}
    P_{m}^{t} = P_{m}^{\mathrm{Fix},t} + P_{m}^{\mathrm{data},t} + P_{m}^{\mathrm{trans},t}.
\end{equation}
As discussed earlier, the fixed power consumption of an \ac{RU} depends solely on its operational mode (active or sleep), and remains constant regardless of traffic load. It can be modeled as: 

\begin{equation}
    P_{m}^{\mathrm{Fix},t} = \alpha_m^t P_m^{\mathrm{active}} + (1 - \alpha_m^t) P_m^{\mathrm{sleep}},
\end{equation}
where $\alpha_m^t \in \{0,1\}$ is the binary activity indicator of \ac{RU} $m$ at time $t$ as we mentioned above. Here, $P_m^{\mathrm{active}}$ and $P_m^{\mathrm{sleep}}$ represent the fixed power consumption levels of \ac{RU} $m$ in active and sleep modes, respectively. Generally, $P_m^{\mathrm{sleep}}$ is significantly lower than $P_m^{\mathrm{active}}$, as many hardware modules are turned off in sleep mode. 

The second component of power consumption is the transmission power, which is proportional to the load of the RU. Specifically, the transmission power of RU $m$ at time slot $t$ is modeled as:

\begin{equation}
    P_{m}^{\mathrm{data},t} = \alpha_m^t \cdot \frac{P_{\mathrm{TX}}}{\eta} \cdot l_m^t = \alpha_m^t \cdot \frac{P_{\mathrm{TX}}}{\eta} \cdot \frac{N_{m}^t}{Q_m},
\end{equation}
where $P_{\mathrm{TX}}$ denotes the maximum transmission power of the RU, and $\eta \in (0,1]$ is the power amplifier (PA) efficiency. Since power amplifiers are not ideal, only a fraction $\eta$ of the consumed electrical power is converted into radiated signal power, and the rest is dissipated as heat. In most case, only half of the consumed power contributes to actual transmission. The power consumption increases with the RU load $l_m^t$, which is defined based on PRB utilization as described in~\eqref{eq:ru_load}. This formulation ensures that higher load levels lead to higher transmission power consumption.

The third component is the transition power, which is incurred when the RU switches its operational state. It is defined as:

\begin{equation}
    P_{m}^{\mathrm{trans},t} = \left| \alpha_m^t - \alpha_m^{t-1} \right| \cdot V_m^{\mathrm{trans}},
\end{equation}

where $\alpha_m^{t-1}$ indicates the state of RU $m$ in the previous time slot, and $V_m^{\mathrm{trans}}$ denotes the energy cost associated with transitioning from sleep mode to active mode. In this work, we consider transition power only when the RU is turned on (i.e., from sleep to active), while mode deactivation (active to sleep) is assumed to incur negligible overhead.

Finally, the total power consumption of the entire network in time slot $t$ is defined as:

\begin{equation}
\begin{aligned}
P_{\text{tot}}^t & = \sum_{m=1}^{M}(P_m^{Fix,t} + P_m^{data,t} + P_{m}^{trans,t})\\
&= \sum_{m=1}^{M}(\alpha_m^t P_m^{active} + (1 - \alpha_m^t) P_m^{sleep}) + \alpha_m^t \frac{P_{TX}}{\eta} * \frac{N_{m}^t}{Q_m} \\
& \quad + ( \left | \alpha_m^t - \alpha_m^{t-1}  \right |*V_m^{trans}).
\end{aligned}
\end{equation}
The objective of network is to minimize the total energy consumption over a given time horizon by dynamically controlling the sleep mode of \acp{RU}, while ensuring \ac{QoS} requirements are satisfied. At the initial time slot ($t = 0$), all RUs are assumed to be active. The optimization problem is formulated as follows:

\begin{equation}
\label{eq:objective function}
\begin{aligned}
\mathcal{P}_1: \quad \min_{\{\alpha_{m}^{t}, n_{m,k}^{t}\}} \quad \sum_{t=1}^{T} & P_{\mathrm{tot}}^t \\
\text{s.t.} \quad & R_{d,k}^t \geq R_{d,k}^{\min}, \quad \forall k \in \mathcal{K},\ \forall t \in \mathcal{T} \\
                  & N_m^t \leq Q_m, \quad \forall m \in \mathcal{M},\ \forall t \in \mathcal{T} \\
                  & \alpha_m^t \in \{0,1\}, \quad \forall m \in \mathcal{M},\ \forall t \in \mathcal{T} \\
                  & \sum_{m \in \mathcal{M}} \delta_{m,k}^t = 1, \quad \forall k \in \mathcal{K},\ \forall t \in \mathcal{T},
\end{aligned}
\end{equation}
where $R_{d,k}^t$ denotes the achieved downlink rate of \ac{UE} $k$ at time $t$, which must satisfy a minimum \ac{QoS} threshold $R_{d,k}^{\min}$. $N_m^t$ is the total number of PRB allocated by RU $m$ and must not exceed its available PRB budget $Q_m$. while $\delta_{m,k}^t \in \{0,1\}$ is an association indicator that ensures that each UE is connected to exactly one RU at any given time. Problem~($\mathcal{P}_1$) aims to minimize the cumulative network energy over a horizon $T$, subject to QoS. Solving this objective with linear or mixed integer formulations in every time interval $t$ would require repeatedly recomputing a large combinatorial program in changing channels and user mobility. Even with relaxations or \ac{MPC}-style solvers, the computational burden grows quickly with the number of RUs and must be incurred at each slot. Moreover, per-slot solutions that optimize instantaneous energy $P_{\text{tot}}^{t}$ cannot anticipate future handovers or load variations, often leading to excessive switching and degraded long-term performance. We therefore cast ($\mathcal{P}_1$) as a \ac{MDP} and learn a policy $\pi(a|s)$ that directly maps the current network state $s$ to RU sleep/active action vector $a$ with the goal of optimizing the long-horizon return. After offline training, online control reduces to a single forward pass of the neural network per slot, which is orders of magnitude lighter than resolving a combinatorial program, while naturally adapting to user mobility since the state already encodes time-varying positions and loads.



\section{DRL-based Energy Saving Mechanisms}

In this section, we develop a centralized \ac{DRL} framework aimed at minimizing the overall energy consumption in O-RAN networks by intelligently controlling the sleep mode of distributed \acp{RU}. Specifically, the proposed approach seeks to determine an optimal activation policy that dynamically switches \acp{RU} between active and sleep modes, while simultaneously guaranteeing the \ac{QoS} requirements of all \acp{UE} in the network.

To the end, we propose three DRL algorithms: \ac{DQNSA}, \ac{DQNMA}, and \ac{TD3}. While DQN-based methods are well-suited to discrete and low-dimensional action spaces, they become less effective in high-dimensional environments with complex combinatorial action structures. In contrast, \ac{TD3}, which extends the DDPG algorithm, offers enhanced performance in continuous control settings through the use of twin critic networks, delayed policy updates, and target smoothing mechanisms. These features make \ac{TD3} particularly suitable for jointly controlling the binary activation states of multiple \acp{RU} in a temporally and spatially dynamic wireless environment.

The remainder of this section first outlines the formal MDP formulation of the RU sleep optimization problem, followed by detailed descriptions of the employed DRL algorithms within the centralized control framework.

\subsection{Markov Decision Process Problem}
In this subsection, we discuss the state space, action space, and reward function of \ac{DRL} model to consist the \ac{MDP} problem.
\subsubsection{State Space} 
state comprises essential information used for policy training. At each time step \( t \), it contains the real data rate of the \acp{UE} , represented as:
\begin{equation}
    \mathbf{R}_{d}^{t} = \begin{bmatrix} R_{d,1}^{t}, R_{d,2}^{t}, \dots, R_{d,K}^{t} \end{bmatrix}^{T},
\end{equation}
which captures the system's ability to meet user demands. The activation states of the radio units (RUs) from the previous time step are denoted as:
\begin{equation}
    \boldsymbol{\alpha}^{t-1} = \begin{bmatrix} \alpha_{1}^{t-1}, \alpha_{2}^{t-1}, \dots, \alpha_{M}^{t-1} \end{bmatrix}^{T}.
\end{equation}
The \ac{PRB} utilization of the RUs is given by:
\begin{equation}
    \mathbf{L}^{t} = \begin{bmatrix} l_{1}^{t}, l_{2}^{t}, \dots, l_{M}^{t} \end{bmatrix}^{T},
\end{equation}
quantifying the load on the RUs. Finally, the spatial positions of the UEs in the network are represented as:
\begin{equation}
    \mathbf{U}^{t} = \begin{bmatrix} (x_{1}^{t}, y_{1}^{t}), (x_{2}^{t}, y_{2}^{t}), \dots, (x_{K}^{t}, y_{K}^{t}) \end{bmatrix}^{T},
\end{equation}
characterized by their \( (x, y) \) coordinates. The \ac{UE} coordinates determine path loss and association and thus affect data rates and RU loads. Including them in $s_t$ enables the agent to capture the spatial distribution and mobility of \acp{UE}. This allows the learned policy to anticipate future handovers, traffic shifts between RUs, and coverage-critical regions, instead of reacting only to instantaneous loads.

In summary, the state can be expressed as:
\begin{equation}
    s_t = \begin{bmatrix} \mathbf{R}_{d}^{t}, \boldsymbol{\alpha}^{t-1}, \mathbf{L}^{t}, \mathbf{U}^{t} \end{bmatrix}^{T}.
\end{equation}

All state parameters are normalized to facilitate a better interpretation by the agent.
\subsubsection{Action Space} 
 action \( \alpha \in \mathcal{A} \) represents the binary operational states of all RUs, defined as:
\begin{equation}
    \mathcal{A} = \left[\alpha_1, \alpha_2, \dots, \alpha_{M}\right].
\end{equation}
Each element \( \alpha_m \) corresponds to an RU, where \( \alpha_m = 1 \) denotes that the RU is active, and \( \alpha_m = 0 \) indicates that the RU is in sleep mode. These actions are derived from the output of the \ac{DRL} models, which guides the energy-efficient operation of the network.
\subsubsection{Reward Function} reward \( r \in \mathbb{R} \) is designed to balance energy efficiency and user satisfaction, guiding the model towards an optimal operational policy. It is defined as:
\begin{equation}
    r = - w_1 \cdot \frac{P_{\text{tot}}}{P_{\text{max}}} - w_2 \cdot \frac{K_{\text{unsat}}}{K},
\end{equation}
where \( P_{\text{max}} \) corresponds to the energy consumption if all \acp{RU} are all active. The term \( K_{\text{unsat}} \) denotes the number of users with data rates below their required thresholds, normalized by the total number of users \( K \). The weights \( w_1 \) and \( w_2 \) adjust the relative importance of energy efficiency and user satisfaction. Normalization is used in the reward function to ensure that the contributions of energy consumption and user satisfaction are scaled to comparable ranges, preventing dominance by one factor over the other. The original RU activation problem Eq \eqref{eq:objective function} can be viewed as minimizing network energy subject to minimum rate constraints. To make the problem tractable within the \ac{MDP} framework, we adopt a penalty-based relaxation in which QoS violations are incorporated into the reward function. This corresponds to a Lagrangian relaxation of the constrained optimization problem, where the penalty weight $w_2$ acts as a Lagrange multiplier. Under sufficiently large $w_2$, the relaxed objective discourages persistent violations and approximates the constrained solution in practice, while maintaining stable gradients for DRL training. During training, multiple weight settings such as $(1,1)$, $(1,3)$, and $(1,10)$ were evaluated. 
The final choice $(1,5)$ was selected because it provides a reasonable balance between energy-saving exploration and QoS protection, avoiding both overly conservative and overly aggressive behavior. We note that this tradeoff is observed under the considered simulation configurations. Overall, At each time slot $t$, the agent observes the state $s_t$, consisting of user data rates, RU loads, RU activation modes, and UE coordinates. Based on this state, the policy $\pi(a_t|s_t)$ outputs an RU activation vector $a_t$. The environment then evolves according to user mobility and traffic dynamics, producing the next state $s_{t+1}$ and the reward $r_t$, which jointly reflect both energy consumption and QoS satisfaction. The agent updates the policy to maximize the long-term cumulative reward, thereby learning optimal \ac{RU} sleep/active decisions.

\subsection{DQN-Based Model Training}

The \ac{DQN} framework extends traditional Q-learning by employing a deep neural network to approximate the action–value function $Q(s,a;\theta)$, where $\theta$ denotes the network parameters. Instead of maintaining a tabular representation, the DQN maps input states to Q-values for all possible actions, enabling decision-making in high-dimensional state spaces. During training, the network parameters are updated to minimize the temporal-difference loss:
\begin{equation}
L(\theta) = \mathbb{E}_{(s,a,r,s') \sim \mathcal{D}} \Big[ \big( r + \gamma \max_{a'} Q(s',a';\theta^{'}) - Q(s,a;\theta) \big)^2 \Big],
\end{equation}
where $\theta'$ is the target network parameter vector, $\gamma$ is the discount factor, $(s,a,r,s')$ are sampled from the replay buffer $D$, and the maximization is taken over the discrete action set. The use of a target network and replay experience helps stabilize training by breaking correlations in sequential samples and preventing oscillations.  

Building on this framework, we applied \ac{DQN} to the \ac{RU} sleep control problem as our initial discrete action solution. In this setting, the agent takes the current network state as input and outputs Q-values for possible RU activation patterns, selecting the action with the highest estimated value. This formulation allows the agent to learn sleep/active scheduling policies without requiring exhaustive optimization at every time step, thus avoiding the computational burden of traditional methods. To investigate how action representation impacts scalability and learning efficiency, we developed two variants: a \ac{DQNMA}, where each action encodes the joint activation state of all \acp{RU}, and a \ac{DQNSA}, where each action controls the state of only one RU at a time. These two designs highlight the trade-off between expressiveness and tractability in DQN-based sleep control.

\subsubsection{Multiple-Action DQN (DQNMA)}

In the DQNMA model, each action represents a complete sleep/active configuration of the entire RU set. Given \( M \) RUs, the total number of possible configurations is \( 2^M \). Instead of representing each configuration as a binary vector, we encode each configuration as an integer index:
\begin{equation}
\mathcal{A}_{\text{multi}} = \{0, 1, \dots, 2^M - 1\}
\end{equation}
Each action \( a \in \mathcal{A}_{\text{multi}} \) is mapped to a binary vector of length \( M \), where the \( i \)-th bit determines whether RU \( i \) is active (\(1\)) or asleep (\(0\)). For example, in a system with \( M = 6 \), the action \( a = 42 \) corresponds to the binary vector \( [1, 0, 1, 0, 1, 0] \), indicating the activation states of all six RUs.

This representation is compact and facilitates Q-value lookup through a single scalar action input. However, the action space still grows exponentially with \( M \), making the Q-table (or output layer of the DQN) increasingly difficult to train. As \( M \) increases, it becomes challenging for the agent to explore all possible configurations effectively, leading to slower convergence and suboptimal learning.

\subsubsection{Single-Action DQN (DQNSA)}

To mitigate the scalability issue inherent in DQNMA, we propose a single-action DQN variant that limits control to one RU per time step. The action space is defined as:
\begin{equation}
    \mathcal{A}_{\text{single}} = \left\{ \text{ON}_1, \dots, \text{ON}_M, \text{OFF}_1, \dots, \text{OFF}_M \right\}
\end{equation}
resulting in a total of \( 2M \) actions. Each action explicitly represents turning ON or OFF a specific RU. This reduces the action space from exponential to linear size, greatly simplifying the learning problem and accelerating convergence. However, the agent's limited per-step control restricts its ability to rapidly adapt to network-wide changes, potentially resulting in suboptimal policies in highly dynamic environments.

\begin{figure}[!t]
	\centering
	\includegraphics[clip, trim=0.0cm 0.cm 0.0cm 0cm, width=1\columnwidth]{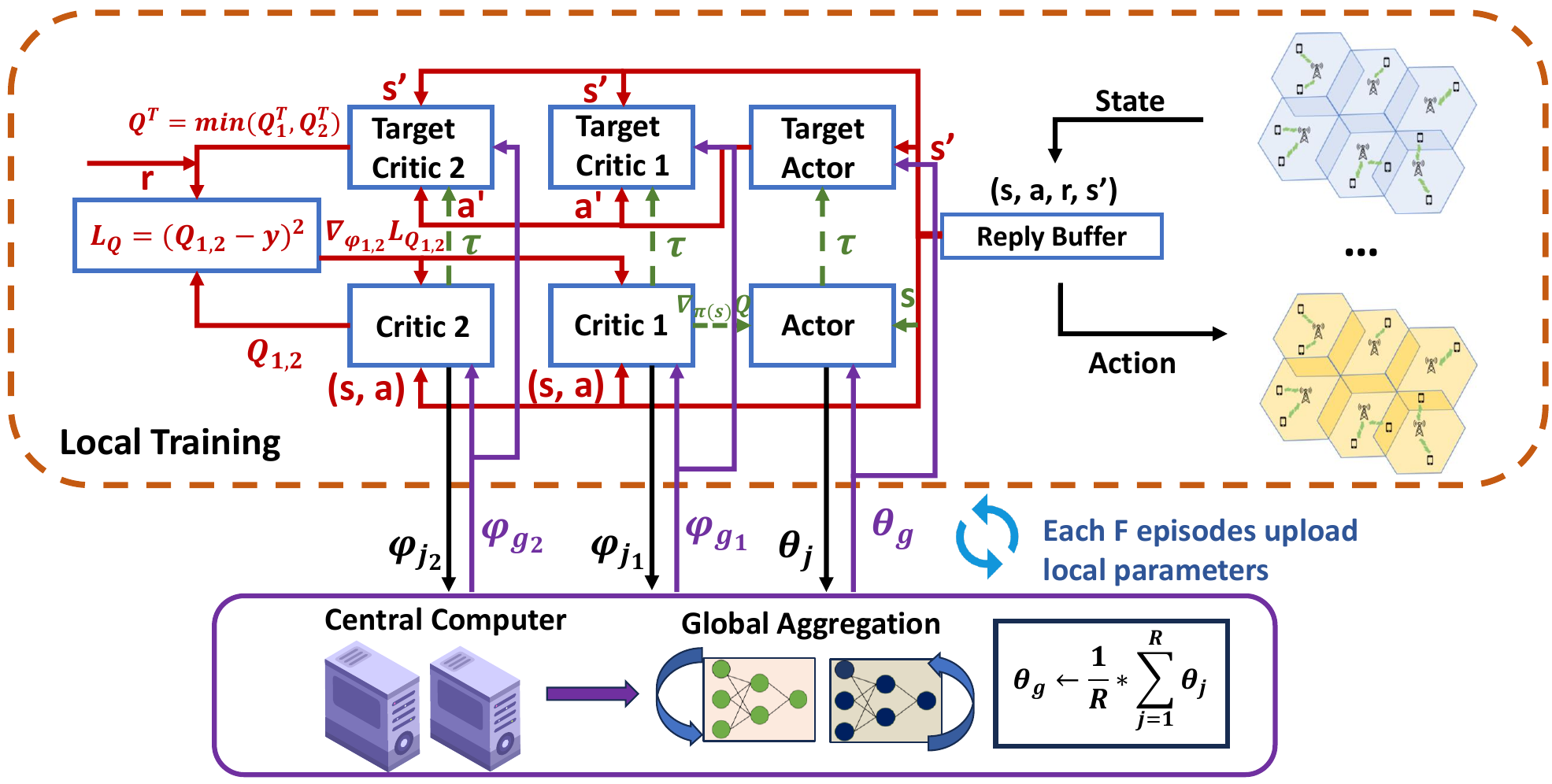}
	\caption{Illustrates the workflow of Fed-TD3 across distributed agents and a aggregator. Red solid lines indicate local critic training based on temporal-difference loss; green dashed lines represent actor updates guided by critic gradients and soft update target network; black and purple solid lines denote the periodic aggregation and redistribution of actor and critic parameters via the aggregator. Each agent operates within its own local environment and contributes to a globally coordinated policy through federated learning.}
	\label{fig:systemmodel}
\end{figure}

\subsection{TD3 Model Training}

Although DQN-based approaches, as we mentioned above, provide effective solutions for discrete RU control, they face significant limitations when applied to large-scale environments with many \acp{RU}. In such scenarios, the action space grows exponentially with the number of controllable \acp{RU}, making it increasingly difficult for value-based methods to explore and learn optimal policies efficiently.
To address these challenges, we adopt a continuous action approach using the \ac{TD3} algorithm as shown in Algorithm \ref{alg:td3}. By modeling the control decisions in a continuous space, \ac{TD3} allows the agent to express more accurate preferences over \ac{RU} activation and significantly reduces the combinatorial complexity of policy learning.

In our adaptation of \ac{TD3}, the actor network outputs a continuous action vector \( \mu_{\theta}(s) \in [0,1]^M \), where each element represents the activation likelihood of an \ac{RU}. To enforce discrete sleep/active decisions in the environment, we apply a deterministic thresholding function with Gaussian exploration noise:
\begin{equation}
a_{i} = f_d(\mu_{\theta}(s)_{i} + \eta_{i}), \quad a_i = \begin{cases}
1 & \text{if } \mu_{\theta}(s)_i + \eta_i > 0.5 \\
0 & \text{otherwise}
\end{cases}
\end{equation}
where, $i\in\{1,\dots,M\}$ indexes the RUs, $\eta_i$ is a Gaussian noise vector with elements $\eta_i\sim\mathcal{N}(0,\sigma^2)$, adding more exploration in the action decision. $f_d(\cdot)$ denotes the discretization mapping from continuous actor outputs to binary RU activation decisions.This approach enables the policy to explore efficiently in a smoothed action space while still producing discrete activation patterns compatible with RU switching.

The environment includes realistic features such as user mobility, handovers, and RU-specific energy models. Transitions \( (s, a, r, s') \) are collected in a replay buffer, and the reward function jointly captures energy efficiency and \ac{QoS} satisfaction. We use two critic networks \( Q_{\phi_1} \) and \( Q_{\phi_2} \) to compute target Q-values and mitigate overestimation bias:

\begin{equation}
    y = r + \gamma \min_{i=1,2} Q_{\phi_i}(s', f_d(\mu_{\theta'}(s') + \epsilon)),
\end{equation}

where \( \epsilon \sim \text{clip}(\mathcal{N}(0, \sigma_{\text{target}}^2)) \) provides target smoothing. The actor is updated using the deterministic policy gradient:

\begin{equation}
\nabla_{\theta} J(\theta) = \mathbb{E}_{s \sim \mathcal{D}}[\nabla_a Q_{\phi}(s, a) \big|_{a = \mu_{\theta}(s)} \cdot \nabla_{\theta} \mu_{\theta}(s)].
\end{equation}

Following TD3’s principle of stability, we delay actor and target network updates relative to the critics, and employ soft target updates:

\begin{equation}
\theta' \leftarrow \tau \theta + (1 - \tau)\theta', \quad \phi_i' \leftarrow \tau \phi_i + (1 - \tau)\phi_i'.
\end{equation}

While our \ac{TD3}-based approach helps the large discrete action space problem by leveraging continuous policy learning and threshold-based execution also provides better results as we show in the simulation section, it does not fully eliminate the scalability limitations of centralized training. As the number of \acp{RU} increases, for example in ultra-dense urban environments, the centralized controller continues to encounter significant challenges related to computational complexity, communication overhead, and limited policy generalization. To address these issues, we further explore a federated \ac{DRL} framework in the next section.

\begin{algorithm}[t]
\small
\caption{TD3-Based Energy-Aware RU Training}
\label{alg:td3}
\KwIn{Initial network state $s_0$, actor parameters $\theta$, critic parameters $\phi_1, \phi_2$, replay buffer $\mathcal{D}$}
\KwOut{Trained actor network $\mu_\theta$ for RU sleep scheduling}
\For{each episode}{
    Initialize environment and receive initial state $s$ \;
    \For{each time step $t = 1, \dots, T$}{
        Add exploration noise $\eta_t \sim \mathcal{N}(0, \sigma^2)$ \;
        Compute continuous action: $a_c = \mu_{\theta}(s) + \eta_t$ \;
        Discretize: $a = f_d(a_c)$ where $a_i = \mathbb{I}[a_{c,i} > 0.5]$\;
        Execute action $a$ in the environment \;
        Receive reward $r$ and next state $s'$ \;
        Store $(s, a, r, s')$ into replay buffer $\mathcal{D}$ \;
        Sample a mini-batch $(s_j, a_j, r_j, s_j')$ from $\mathcal{D}$ \;
        Compute target action: $\tilde{a}_j' = f_d(\mu_{\theta'}(s_j') + \epsilon)$ \;
        Compute target Q-value: $y_j = r_j + \gamma \min_{i=1,2} Q_{\phi_i'}(s_j', \tilde{a}_j')$ \;
        Update critics $\phi_1, \phi_2$ by minimizing:
        $\mathcal{L}(\phi_i) = \frac{1}{N} \sum_j \left( Q_{\phi_i}(s_j, a_j) - y_j \right)^2$ \;
        \If{every $d$ steps}{
            Update actor using deterministic policy gradient \;
            Soft-update target networks: $\phi_i' \leftarrow \tau \phi_i + (1-\tau)\phi_i'$, $\theta' \leftarrow \tau \theta + (1-\tau)\theta'$ \;
        }
        $s \leftarrow s'$
    }
}
\end{algorithm}

\begin{algorithm}[ht]
\scriptsize
\caption{Federated TD3 Training Process}
\label{alg:FedTD3}
\begin{algorithmic}[1]
\REQUIRE Multi-region environment with $R$ regions, local TD3 agents $\{A_i\}_{i=1}^{R}$, aggregation frequency $F$ and TD3 hyperparameters.
\STATE \textbf{Initialization:}
    \FOR{$i=1$ \TO $R$}
        \STATE Initialize local agent $A_i$ with its actor, critics, and target networks.
    \ENDFOR
\STATE Initialize the global model parameters $\theta_{\text{global}}$, $\phi_{\text{global\_1}}$ and $\phi_{\text{global\_2}}$ with random value. Establish the connection between local agents and a global server.
\STATE \textbf{Training Loop:}
\FOR{$\text{episode}=1$ \TO $N_{eps}$}
    \FOR{$j=1$ \TO $R$}
        \STATE Initialize state $s_0^{(i)}$.
    \FOR{$t=0$ \TO $T-1$}
            \STATE Agent $A_i$ selects an action $a_t^{(i)}$
            \STATE Execute actions $a_t^{(i)}$ in the environment.
            \STATE Obtain next states $s_{t+1}^{(i)}$ and rewards $r_t^{(i)}$.
            \STATE Agent $A_i$ stores transition $(s_t^{(i)},a_t^{(i)},r_t^{(i)},s_{t+1}^{(i)})$.
            \STATE Agent $A_i$ performs a local TD3 update.
            \IF{$(t+1) \bmod F = 0$}
                \STATE Global Server Aggregation (FedAvg). Agent $A_i$ sends $\theta_{i}$, $\phi_{i\text{\_1}}$ and $\phi_{i\text{\_2}}$ to the global server.
        \ENDIF
        \STATE Update state: $s_t^{(i)} \gets s_{t+1}^{(i)}$.
        \ENDFOR
    \ENDFOR
\ENDFOR
\STATE \textbf{Global Server Aggregation (FedAvg):}
\STATE the The global server receives local parameters from all agents.
\STATE compute the weighted average of actor and critic model update: \\
$\theta_{\text{avg}} = \sum_{j}(\omega_{j} \cdot \theta_{j} / \sum_{j}\omega_{j})$\\
$\phi_{\text{avg}\_i} = \sum_{j}(\omega_{j\_i} \cdot \phi_{j\_i} / \sum_{j\_i}\omega_{j\_i}),\quad i=1,2.$ Where $\omega_{i}$ are weight factors
\STATE \textbf{Global Model Update:}
\STATE Update the global actor and critic model parameters:\\
$\theta_{\text{global}} = \theta_{\text{avg}} $, \quad
$\phi_{\text{global}\_i} = \phi_{\text{avg}\_i}, \quad i=1,2$,\\
$\theta_{\text{global}}'=\theta_{\text{global}}$,\quad
$\phi_{\text{global}\_i}' = \phi_{\text{global}\_i},\quad i=1,2$.
\STATE \textbf{Global Distribution:} For each region $j$, update local parameters: $ \theta_{j} \gets \theta_{\text{global}}$, $\phi_{j\_i} \gets \phi_{\text{global}\_i}  \quad i=1,2$,\\
$\theta_{j}' \gets \theta_{\text{global}}'$, $\phi_{j\_i}' \gets \phi_{\text{global}\_i}'$.
\end{algorithmic}
\end{algorithm}

\section{Federated DRL for Scalable RU Control in O-RAN}

The modular and disaggregated architecture of O-RAN makes it a natural fit for federated \ac{DRL}, particularly in scenarios that involve large-scale and geographically distributed RU deployments. In O-RAN, control functionality is split across multiple layers, such as xApps operating at the Near-RT RIC and rApps at the Non-RT RIC, communicating via standardized open interfaces. This structure aligns seamlessly with the federated learning paradigm, where learning agents (e.g., xApps) act as local clients that train policies independently using site-specific data and periodically synchronize with a central coordinator (e.g., an rApp) to build a global policy model \cite{ndikumana2023federated,wang2024hierarchical}.

Adopting federated \ac{DRL} in this context offers significant scalability advantages over centralized DRL. As network size and RU density increase, centralized learning suffers from exploding state and action spaces, as well as communication bottlenecks and slow convergence. By distributing the learning process across O-RAN components, federated \ac{DRL} reduces computational and communication loads at any single point \cite{lim2020federated,singh2024communication}, while also enabling site-level policy customization. Moreover, the reuse of O-RAN’s open interfaces allows efficient and standards-compliant integration of FL workflows into real network deployments. An overview of the proposed federated DRL framework within the O-RAN architecture is depicted in Fig.~\ref{fig:systemmodel1}. In this design, each O-RAN region is equipped with a Near-RT RIC hosting local DRL agents that interact with the underlying RUs and UEs to make real-time sleep/active decisions. These agents are trained locally using region-specific traffic and mobility dynamics, thereby capturing heterogeneous environmental characteristics without requiring raw data exchange. Periodically, the locally trained models are uploaded to the Non-RT RIC, which serves as the global aggregator. The Non-RT RIC integrates model updates from multiple regions to construct a global policy, which is then redistributed back to the Near-RT RICs. This hierarchical workflow aligns naturally with the functional split of O-RAN: the Near-RT RIC ensures responsiveness to fast-varying network states, while the Non-RT RIC provides global coordination, scalability, and policy generalization across distributed deployments.

\subsection{Federated DRL Training and Global Aggregation}

In our federated DRL framework, each agent corresponds to a distinct geographical region, consisting of multiple \acp{RU} and \acp{UE}. The training process begins with the aggregator distributing an initialized global model to all participating agents. This model includes either Q-network weights for DQN agents or actor–critic parameters for TD3 agents, depending on the algorithm. Fig. \ref{fig:systemmodel} illustrates the entire Fed-TD3 training process as a representation.
After receiving the global model, each agent independently interacts with its local environment, collecting state–action–reward transitions and updating its DRL model based on region-specific user mobility and traffic dynamics. This local training proceeds over a fixed number of episodes $F$.
Once the local training phase is completed, each agent uploads selected model parameters to the aggregator. The aggregator then performs model aggregation to produce an updated global model. This aggregation strategy varies depending on the type of DRL model used and will be detailed in the following section.
The updated global model is subsequently redistributed to all agents for the next round of training. This iterative cycle of local update and global synchronization continues until convergence. By leveraging this decentralized optimization strategy, our framework ensures that agents collaboratively learn a generalizable policy while preserving privacy and significantly reducing communication overhead.

Depending on the underlying DRL algorithm, we employ two distinct parameter-sharing strategies to support both value-based and policy-based learning. The complete Fed-TD3 training process is summarized in Algorithm~\ref{alg:FedTD3}.
For agents using the DQN architecture with discrete action spaces, each agent performs Q-learning on local environment and periodically uploads its Q-network to the server. The global model is computed by averaging the Q-values across all agents:
\begin{equation}
Q_{\text{global}}(s,a) = \frac{1}{R} \sum_{j=1}^{R} Q_j(s,a), \quad \forall s,a,
\end{equation}
\begin{equation}
Q_j(s,a) \leftarrow Q_{\text{global}}(s,a), \quad \forall s,a,
\end{equation}
where \( Q_j(s,a) \) represents the Q-network of agent \( j \), and \( R \) is the number of agents. The aggregated global model \( Q_{\text{global}}(s,a) \) is then broadcast to all agents for subsequent training rounds.

For agents using the TD3 algorithm in the continuous action space, each agent maintains an actor–critic architecture. The global policy seeks to learn $|\mathcal{S}| \times|\mathcal{A}|$ table, $\pi_{global}(a|s)$. After several local update episodes, the policy network which is the actor network parameters is uploaded to the server for aggregation:
\begin{equation}
\pi_{\text{global}}(a|s) = \frac{1}{R} \sum_{j=1}^{R} \pi_j(a|s), \quad \forall s,a,
\end{equation}
\begin{equation}
\pi_j(a|s) \leftarrow \pi_{\text{global}}(a|s), \quad \forall s,a.
\end{equation}

Although the standard actor–critic framework typically aggregates only the actor to ensure decentralized critic adaptation, in our design we also upload and aggregate the two critic networks across agents. This additional aggregation improves training stability and enhances the generalizability of the learned value functions. After each global aggregation step, both the actor and critic target networks are updated, as detailed in Algorithm~\ref{alg:FedTD3}. In our scenario, the areas share a homogeneous RU activation task and similar traffic patterns, which reduces the degree of heterogeneity. In the future works we will investigate heterogeneous environments that cause further challenges. In this work we mainly want to show that Federated learning is a scalable AI solution for O-RAN networks.

\section{Simulation Results}
To evaluate the effectiveness of the proposed energy savings framework, we perform simulations across a range of network scenarios and baseline models. Our evaluation includes centralized and federated learning environments to assess performance in terms of reward, energy consumption, and scalability.

\begin{figure}[!t]
	\centering
	\includegraphics[clip, trim=0.0cm 0.cm 0.0cm 0cm, width=1\columnwidth]{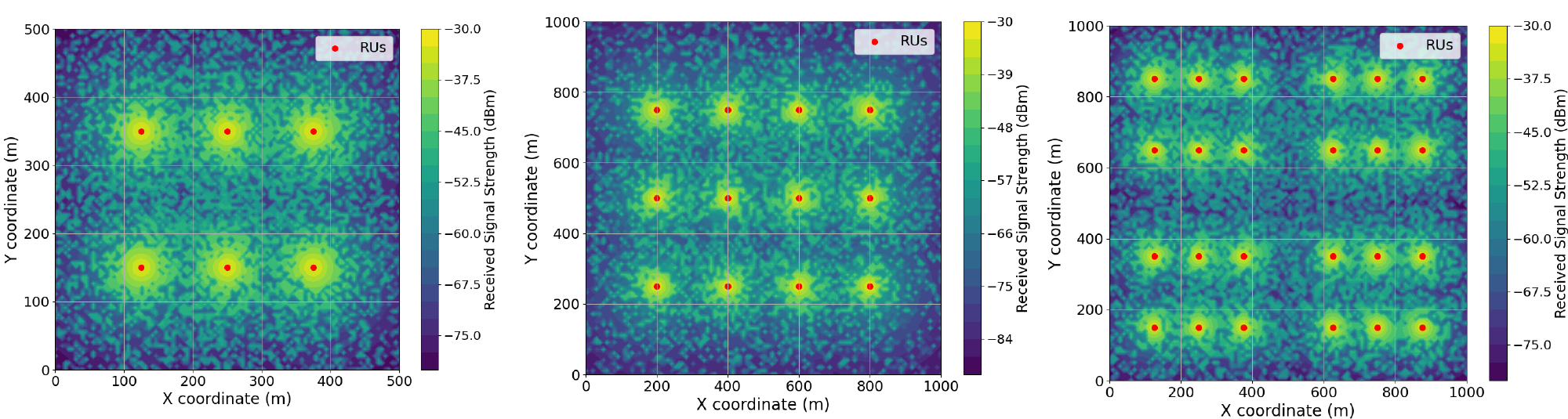}
	\caption{The left plot shows a radio map of single 500\,m$\times$500\,m area used for centralized  training and evaluation. 
    The right plot depicts a 1000\,m$\times$1000\,m area composed of four such subregions, 
    representing a composite environment for centralized training and inference. 
    This comparison setting is used to evaluate the generalization capability of the global model trained via federated reinforcement learning versus a centralized model trained on the combined area. The middle plot presents a single large-scale 1000\,m$\times$1000\,m environment used to evaluate model performance under centralized DRL training. }
        \label{fig.radiomap}
\end{figure}

\subsection{Simulation Settings}
In our simulations, each scenario is modeled as an \ac{O-RAN} environment where $M$ \acp{RU} serve $K$ \acp{UE} within a square area of size $L \times L$\,m$^2$. In practical O-RAN deployments, the Near-RT RIC is commonly deployed in edge cloud or regional edge data centers with latency budgets between 10\,ms and 1\,s. In our system model, the inference of the DRL agent runs inside the Near-RT RIC. The model training or federated aggregation may take place at the same edge location. This placement ensures that the decision-making latency requirements of the xApp are satisfied. As illustrated in Fig.~\ref{fig.radiomap}, three layouts are considered: 
(i) a $500$\,m$\times500$\,m area with 6~\acp{RU}, used both as the centralized DRL testbed and as the local region in \ac{FL}; 
(ii) a $1000$\,m$\times1000$\,m area with 12~\acp{RU}, representing a large-scale single region; 
(iii) a $1000$\,m$\times1000$\,m composite area with 24~\acp{RU}, formed by four independent $500$\,m$\times500$\,m subregions, used to examine the scalability of \ac{FL}.

The \acp{UE} move at a constant speed $v = v_{\text{avg}} \pm v_{\text{std}}$ (mean $v_{\text{avg}}=2$\,m/s, standard deviation $v_{\text{std}}=0.5$\,m/s), with individual speeds randomly drawn from this range. A periodic mobility pattern is applied, where \acp{UE} travel from the service area edge toward the center and back, forming a structured cyclic movement. 
The wireless channel adopts the 3GPP \ac{UMi} model \cite{3gppTR38901} with both \ac{LOS} and \ac{NLOS} conditions. For LOS links:  
\begin{equation}
\begin{aligned}
PL_{m,k}^{\text{LOS}} =
\begin{cases}
32.4 + 21 \log_{10}(d_{m,k}) \\
\quad {}+ 20 \log_{10}(f), & d_{m,k} \le d_{\text{BP}}, \\[4pt]
32.4 + 40 \log_{10}(d_{m,k}) \\
\quad {}+ 20 \log_{10}(f) - 9.5, & d_{m,k} > d_{\text{BP}} ,
\end{cases}
\end{aligned}
\end{equation}
with $d_{\text{BP}} = \frac{4 h_{\text{RU}} h_{\text{UE}} f}{c}$ as the breakpoint distance. For NLOS links:  
\begin{equation}
\begin{aligned}
PL_{m,k}^{\mathrm{NLOS}} =
&\, 35.3 + 22.4 \log_{10}(d_{m,k}) \\
&\, {}+ 21.3 \log_{10}(f)
      - 0.3\bigl(h_{\mathrm{UE}} - 1.5\bigr).
\end{aligned}
\end{equation}

The LOS probability is:
\begin{equation}
P_{\text{LOS}} = \min \left( \frac{18}{d_{m,k}}, 1 \right) \left(1 - e^{-d_{m,k}/36} \right) + e^{-d_{m,k}/36},
\end{equation}
and $P_{\text{NLOS}} = 1 - P_{\text{LOS}}$, with $P_{\text{LOS}}=1$ if $d_{m,k}<18$\,m.
Other system parameters, including carrier frequency, antenna heights, bandwidth, and RU power consumption values, follow Table~\ref{tab:sim_params}.

\begin{table}[h]
\centering
\small
\caption{Simulation Parameters}
\label{tab:sim_params}
\begin{tabular}{|l|c|}
\hline
\textbf{Parameter} & \textbf{Value} \\ \hline
TD3 Actor learning rate ($\alpha_\pi$) & 0.0001 \\ \hline
TD3 Critic learning rate ($\alpha_Q$) & 0.001 \\ \hline
DQN learning rate ($\alpha$) & 0.0001 \\ \hline
Mini-batch size ($B$) & 128 \\ \hline
Replay memory size ($\mathcal{R}$) & 50000 \\ \hline
Discount factor ($\gamma$) & 0.99 \\ \hline
Training episodes ($N_{\text{eps}}$) & 2000 \\ \hline
Random seed & 42 \\ \hline
Soft update coefficient ($\tau$) & 0.01 \\ \hline
Carrier frequency ($f$) & 2 GHz \\ \hline
RU height ($h_{\text{RU}}$) & 15 m \\ \hline
UE height ($h_{\text{UE}}$) & 1.7 m \\ \hline
Network size ($L$) & {[}500, 1000{]} m \\ \hline
Number of RUs ($M$) & {[}6, 12, 24{]} \\ \hline
Number of UEs ($K$) & {[}20--80{]} \\ \hline
Number of local regions ($R$) & 8 \\ \hline
Minimum data rate requirement ($R_{\text{min}}$) & 3 Mbps \\ \hline
Noise power ($\sigma_n^2$) & -174 dBm/Hz \\ \hline
Power amplifier efficiency ($\eta$) & 0.5 \\ \hline
Average UE speed ($v_{\text{avg}}$) & 2 m/s \\ \hline
Std. deviation of UE speed ($v_{\text{std}}$) & 0.5 m/s \\ \hline
Active mode RU power ($P^{\text{active}}$) & 20 W \\ \hline
Sleep mode RU power ($P^{\text{sleep}}$) & 5 W \\ \hline
Maximum transmission power ($P_{\text{TX}}$) & 1 W / 30 dBm \\ \hline
Mode transition power ($P^{\text{trans}}$) & 3 W \\ \hline
Reward weights $(w_1, w_2)$ & 1, 5 \\ \hline
\end{tabular}
\end{table}

During training, each episode consists of 200 time steps. Considering the transition latency of \acp{RU}, each time step is set to 1\,s, which is also the interval for both energy consumption measurement and sleep/active mode decisions. The energy consumption recorded at each step corresponds to the instantaneous value, while the total energy consumption is obtained by summing over all steps in an episode. For the \ac{TD3} model, both the actor and critic networks consist of four fully connected layers, with activation functions and layer sizes detailed in Table~\ref{tab:network_architecture}. The Adam optimizer is used for parameter updates, and \ac{BN} is applied to improve training stability. For the DQN-based models, each network contains five fully connected layers, with architectures also provided in Table~\ref{tab:network_architecture}.

We evaluate three centralized DRL approaches: \ac{DQNSA}, \ac{DQNMA}, and \ac{TD3}, which together provide a representative coverage of both value-based and policy-based methods. To investigate scalability in large-scale or distributed deployments, each of these approaches is further extended to a federated learning setting, resulting in the Fed-DQNSA, Fed-DQNMA, and Fed-TD3 models. In the federated setting, agents are trained on separate local regions and periodically aggregated into a global model, whereas the centralized setting relies on a single combined environment. This design enables a systematic comparison between centralized and federated paradigms across different DRL architectures.

\begin{table}[h]
\centering
\caption{Network Configurations}
\resizebox{\columnwidth}{!}{
\label{tab:network_architecture}
\begin{tabular}{|l|l|l|l|l|l|}
\hline
                & \textbf{Layer1} & \textbf{Layer2} & \textbf{Layer3} & \textbf{Layer4} & \textbf{Output Layer} \\ \hline
\textbf{Actor}  & BN+relu 512     & relu 256        & relu 128        & None            & sigmoid $M$             \\ \hline
\textbf{Critic} & relu 512        & relu 256        & relu 128        & None            & linear 1              \\ \hline
\textbf{DQNSA}  & relu 512        & relu 384        & relu 256        & relu 128        & linear $M^2$             \\ \hline
\textbf{DQNMA}  & relu 512        & relu 384        & relu 256        & relu 128        & linear $2M$             \\ \hline
\end{tabular}
}
\end{table}

\subsection{Numerical Results}
\begin{figure}[!t]
	\centering
	\includegraphics[clip, trim=0.0cm 0.cm 0.0cm 0cm, width=1\columnwidth]{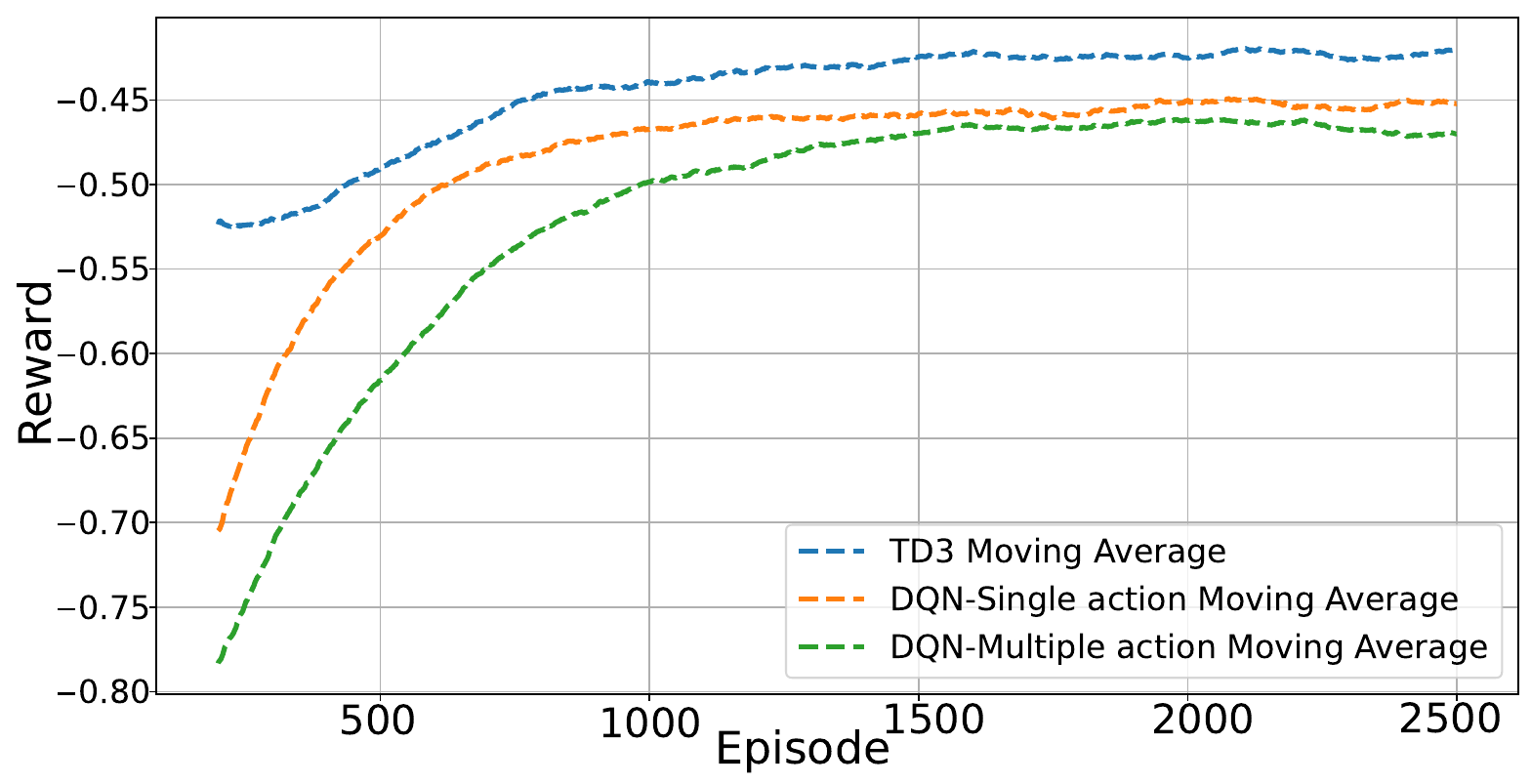}
	\caption{Illustrates the rewards of TD3, DQNMA and DQNSA model in 500\,m$\times$500\,m area with 6 RUs and 20 UEs.}
    \label{fig:3models-500-reward}
\end{figure}

\begin{figure}[!t]
	\centering
	\includegraphics[clip, trim=0.0cm 0.cm 0.0cm 0cm, width=1\columnwidth]{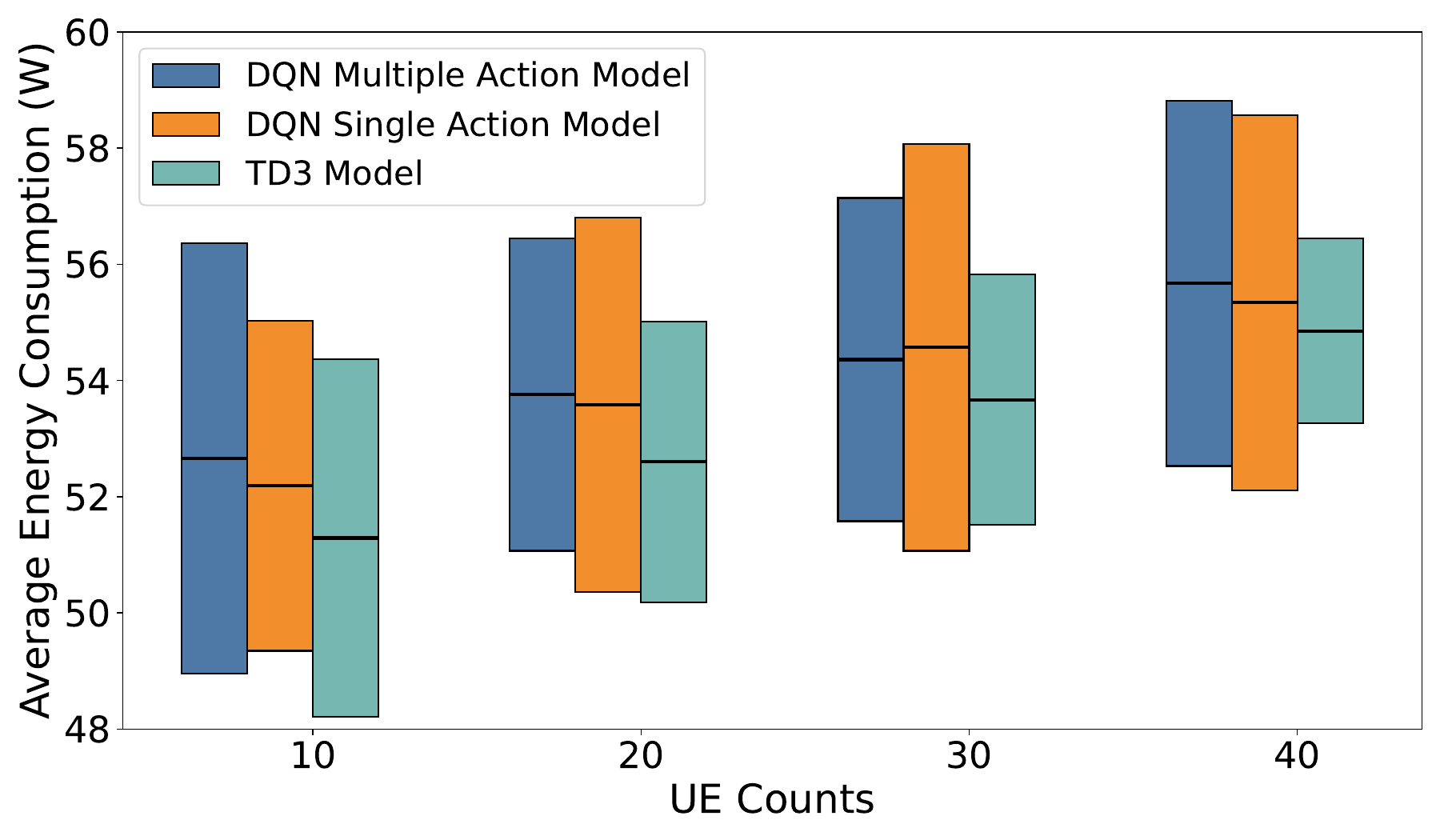}
	\caption{Illustrates the average energy consumption in 500\,m$\times$500\,m area among 6 RUs and 10 to 40 \acp{UE} with DQNMA, DQNSA and TD3 models. The maximum theoretical energy consumption is 126W.}
    \label{fig:3models-500-power}
\end{figure}

In Fig.~\ref{fig:3models-500-reward}, we present the reward performance of the centralized \ac{TD3}, \ac{DQNSA}, and \ac{DQNMA} models over 2500 episodes in a 500,m $\times$ 500,m network layout. The results clearly highlight the advantages of continuous-action methods: the TD3 model consistently outperforms both DQN variants, achieving the highest long-term reward with smooth convergence and superior stability. In contrast, the \ac{DQNSA} model demonstrates moderate performance, yet its reward trajectory exhibits noticeable fluctuations, reflecting the limitations of discrete-action value approximation in adapting to dynamic environments. The \ac{DQNMA} model, while extending the action granularity with multiple discrete outputs, does not translate this added complexity into performance gains; instead, it yields the lowest reward among the three. This comparison indicates that simply enlarging the discrete action space within a DQN framework cannot effectively capture the nuanced trade-offs required for energy-efficient RU scheduling, whereas TD3’s continuous-action formulation enables more fine-grained and adaptive decision-making.

Fig.\ref{fig:3models-500-power} indicates the energy consumption performance in 500\,m$\times$500\,m area for varying \ac{UE} counts ranging from 10 to 40, is still compared among the previous three models. The figure distinctly shows the average energy consumption, indicated by the black line in the center of each bar, while the length of each bar represents the variance in energy consumption. With a total of 6 \acp{RU} installed in this area, the maximum possible energy consumption reaches 126W when all \acp{RU} operate in active mode. The results clearly demonstrate that all three models achieve significant energy savings, exceeding 50\% compared to the theoretical maximum. Specifically, the \ac{TD3} model achieving up to an additional 6\% energy saving compared to the two \ac{DQN}-based models.

\begin{figure}[!t]
	\centering
	\includegraphics[clip, trim=0.0cm 0.cm 0.0cm 0cm, width=1\columnwidth]{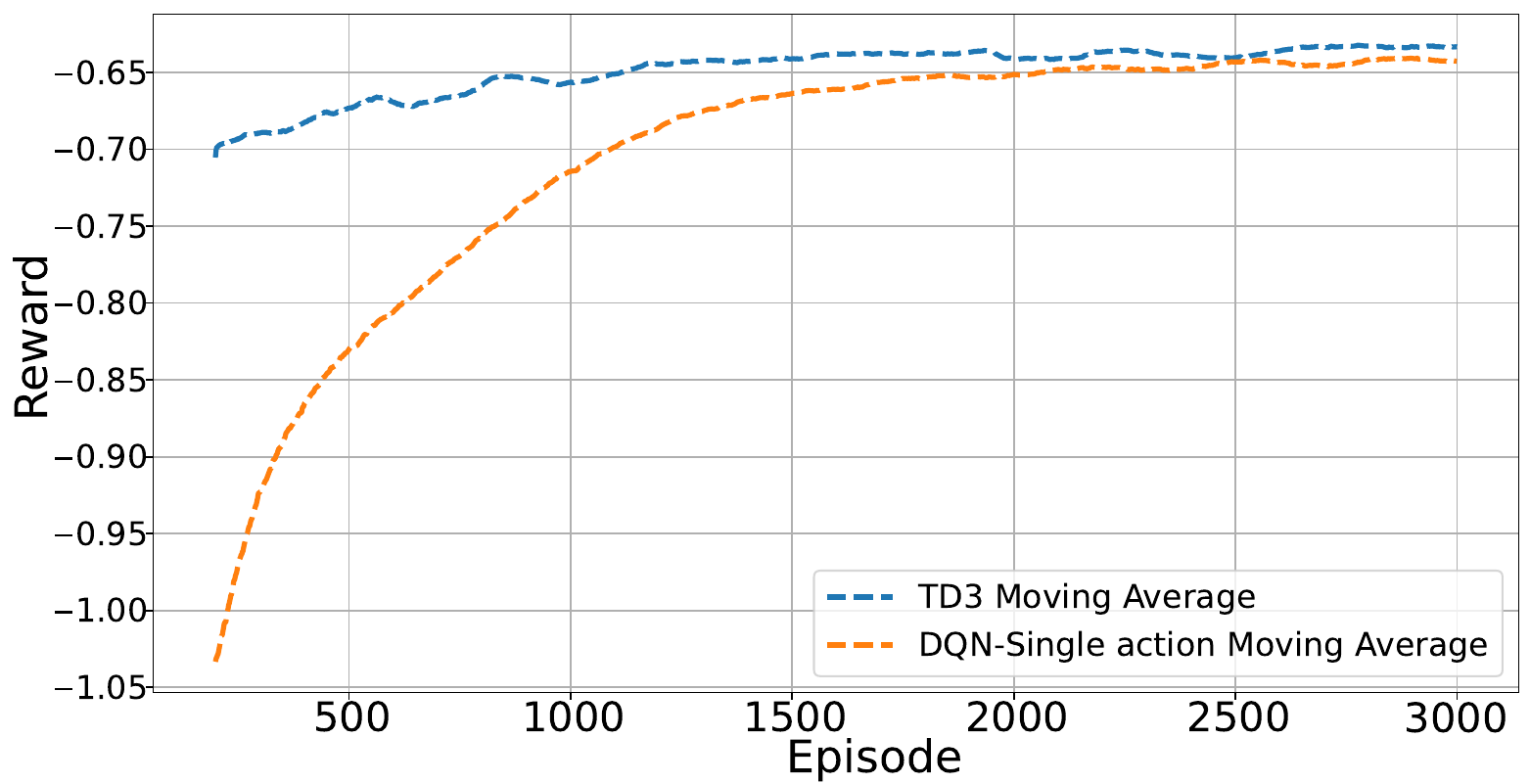}
	\caption{Illustrates the rewards of TD3 and DQNSA model in 1000\,m$\times$1000\,m area with 12 RUs and 40 UEs.}
        \label{fig:2models-1000-reward}
\end{figure}

\begin{figure}[!t]
	\centering
	\includegraphics[clip, trim=0.0cm 0.cm 0.0cm 0cm, width=1\columnwidth]{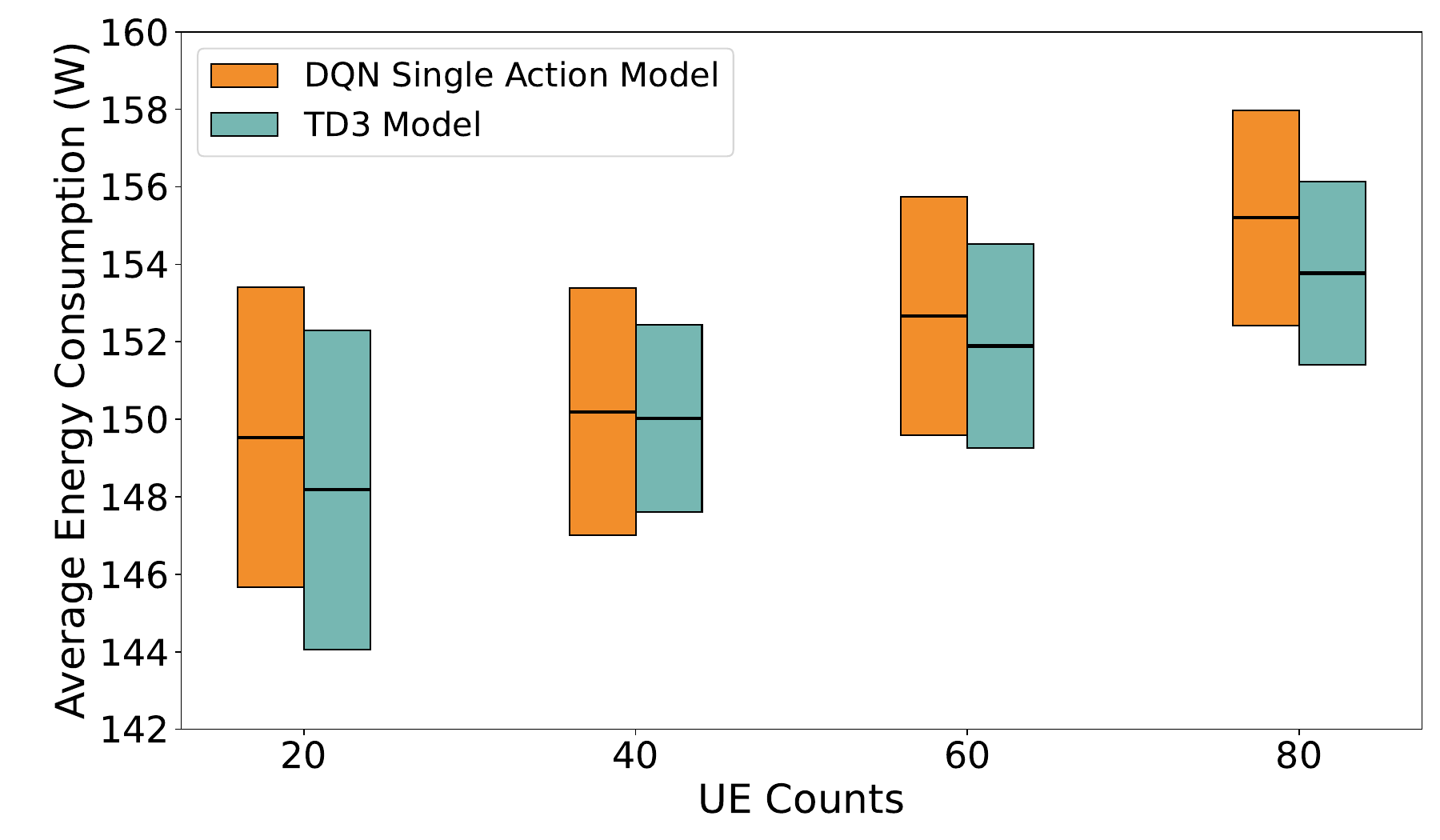}
	\caption{Illustrates the average energy consumption among 12 RUs and 20 to 80 \acp{UE} in 1000\,m$\times$1000\,m area with TD3 and DQNSA models. The maximum theoretical energy consumption is 252W.}
        \label{fig:2models-1000-power}
\end{figure}

In Fig.\ref{fig:2models-1000-reward} shows the reward performance over an extended simulation environment from 500\,m $\times$500\,m area to 1000\,m$\times$1000\,m area. But only compare the \ac{TD3} model against the \ac{DQNSA} model due to the excessive expansion of \ac{DQNMA} action space (e.g. 2$^{12}$). In the results, the \ac{TD3} model still achieves higher and stable reward levels throughout all episodes. Furthermore, the TD3 model demonstrates faster and more efficient convergence toward optimal solutions. The Fig.\ref{fig:2models-1000-power} illustrates the energy consumption between the \ac{TD3} and \ac{DQNSA} model, covering \ac{UE} counts from 20 to 80. Numerically, the \ac{TD3} model maintains lower energy consumption, approximately 144w at 20 UEs and up to around 152w at 80 UEs, significantly below the theoretical maximum consumption of 252W (more than 40\%) with all 12 RUs active. However, the TD3 model can also achieve 5\% over the \ac{DQNSA} mode, further emphasizing the TD3 model in a complex and large-scale environment.

\begin{figure}[!t]
	\centering
	\includegraphics[clip, trim=0.0cm 0.cm 0.0cm 0cm, width=1\columnwidth]{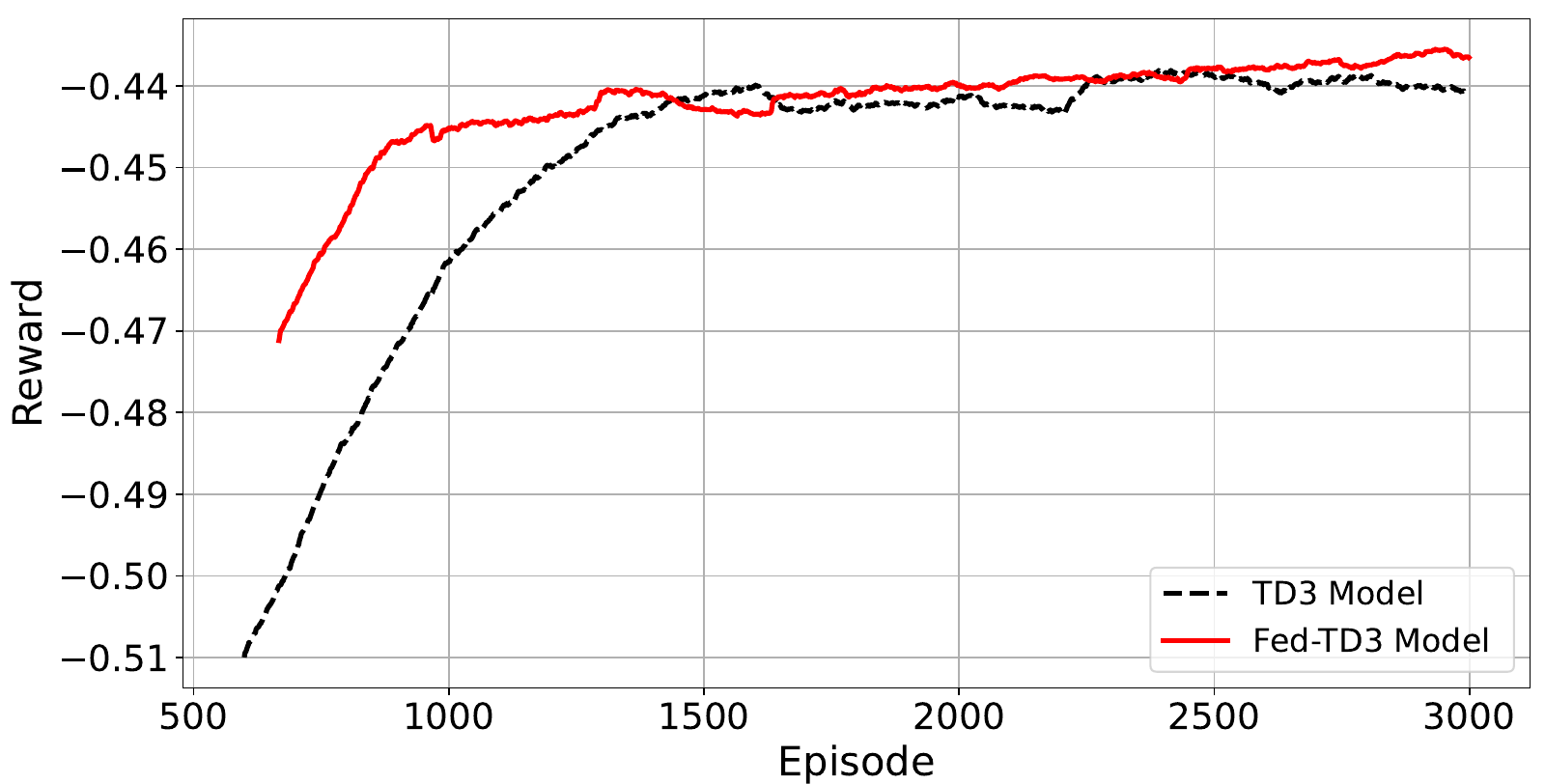}
	\caption{Illustrates the rewards and convergence speed of TD3 and Fed-TD3 in 500\,m$\times$500 \,m area with 6 RUs and 20 UEs.}
        \label{fig:Fed-TD3&TD3}
\end{figure}
Having established that TD3 provides the best performance among the centralized DRL approaches, we next extend our analysis to examine the benefits of federated learning. As illustrated in Fig.~\ref{fig:Fed-TD3&TD3}, the training reward curves highlight clear advantages of the federated framework. To statistically validate the convergence behavior, we computed the convergence episode for five independent runs of both TD3 and Fed-TD3 and performed a Welch two-sample t-test. Fed-TD3 converges in 
$858.8 \pm 127.4$ episodes on average, while TD3 requires 
$1421.4 \pm 322.6$ episodes. Fed-TD3 achieves 40\% reduction in convergence episodes compared with TD3. The t-test yields $p = 0.0309 < 0.05$, indicating that the faster convergence of Fed-TD3 
is statistically significant at the 95\% confidence level. The observed behavior is interpreted as an empirical outcome of the proposed multi-area hierarchical learning setup. The improved learning dynamics can be attributed to the aggregation of diverse policy updates across regions, which enhances generalization and stabilizes the learning process. Moreover, Fed-TD3 attains a higher final reward compared to centralized TD3, indicating superior long-term performance in balancing energy savings and service quality.

\begin{figure}[!t]
	\centering
	\includegraphics[clip, trim=0.0cm 0.cm 0.0cm 0cm, width=1\columnwidth]{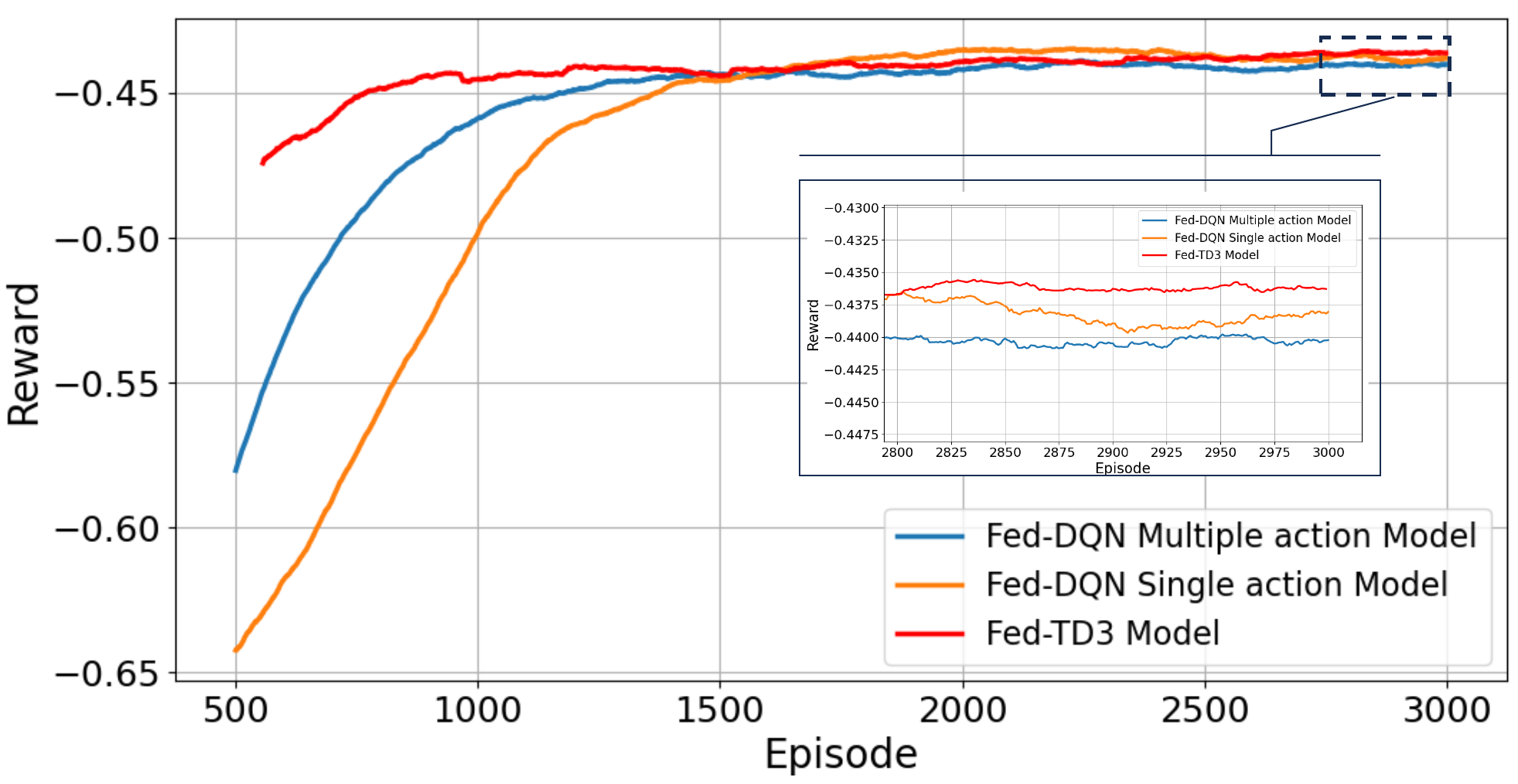}
	\caption{Illustrates the rewards and convergence speed of Fed-TD3, Fed-DQNSA and Fed-DQNMA model in 500\,m$\times$500\,m area with 6RUs and 20UEs.}
        \label{fig:Fed-TD3&Fed-DQN}
\end{figure}

Fig. \ref{fig:Fed-TD3&Fed-DQN} compares the training reward curves of three federated DRL models: Fed-TD3, Fed-DQNMA, and Fed-DQNSA. Both federated DQN variants exhibit significant improvements over their non-federated counterparts, with approximately a 10\% increase in final average reward. This demonstrates that federated learning effectively enhances the generalization and learning efficiency of value-based methods even under discrete action settings. Notably, the final rewards achieved by Fed-DQNMA and Fed-DQNSA closely approach that of Fed-TD3, suggesting that with appropriate aggregation, discrete-action methods can benefit substantially from distributed training.
However, the convergence speed of Fed-DQNMA and Fed-DQNSA remains noticeably slower than Fed-TD3. While Fed-TD3 stabilizes after approximately 900 episodes, the federated DQN models require around 1,300 episodes to reach similar stability, indicating a relative slowdown of about 30\%. This gap reflects the inherent limitations of Q-learning in handling large action spaces compared to policy-based approaches.

\begin{figure}[!t]
    \centering
    \begin{subfigure}[b]{0.45\textwidth}
        \centering
        \includegraphics[width=1\textwidth]{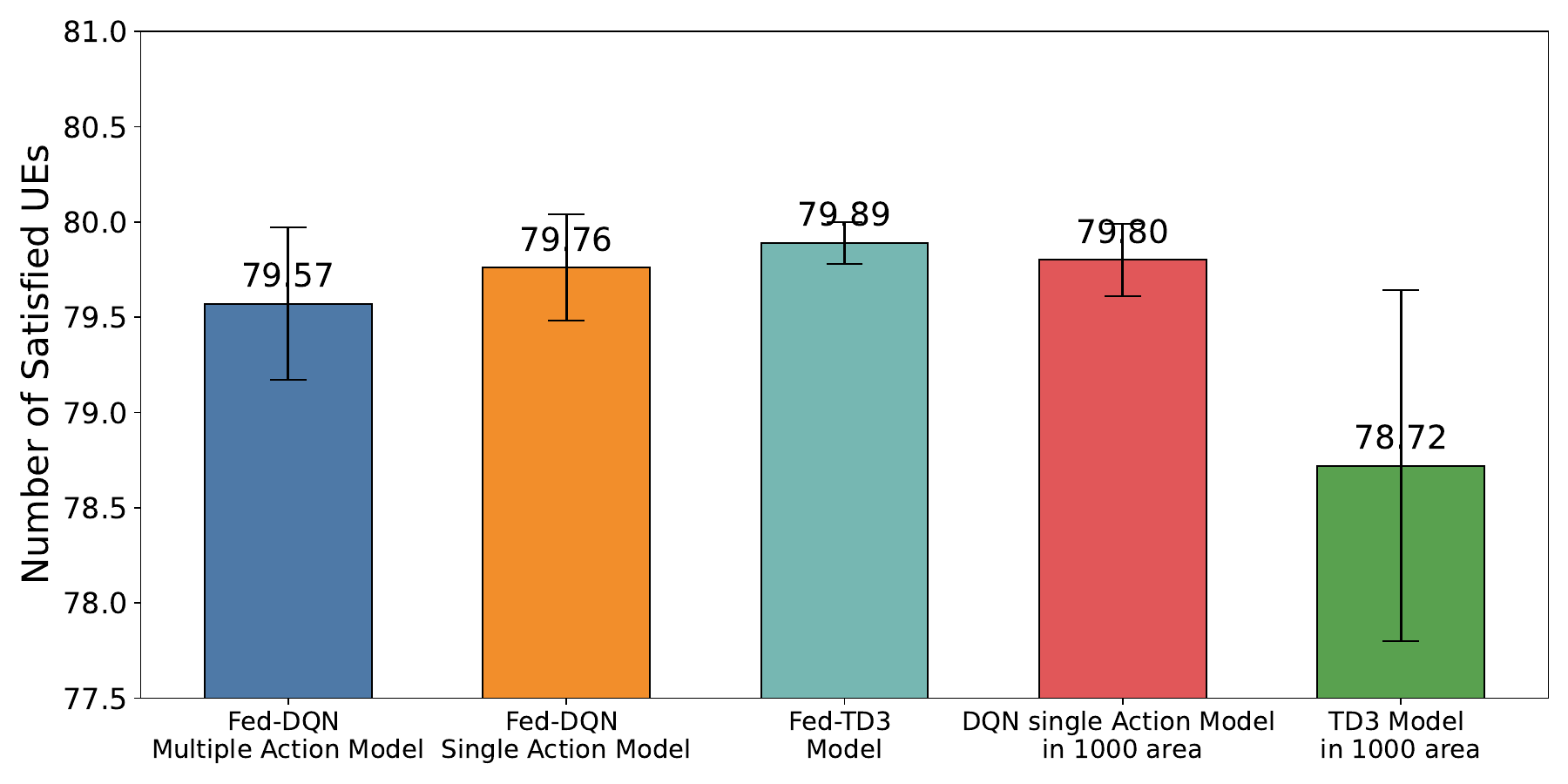}
        \label{fig:5models_UE_QoS}
    \end{subfigure}
    \hfill
    \begin{subfigure}[b]{0.45\textwidth}
        \centering
        \includegraphics[clip, trim=0.0cm 0.cm 0.0cm 0cm, width=1\textwidth]{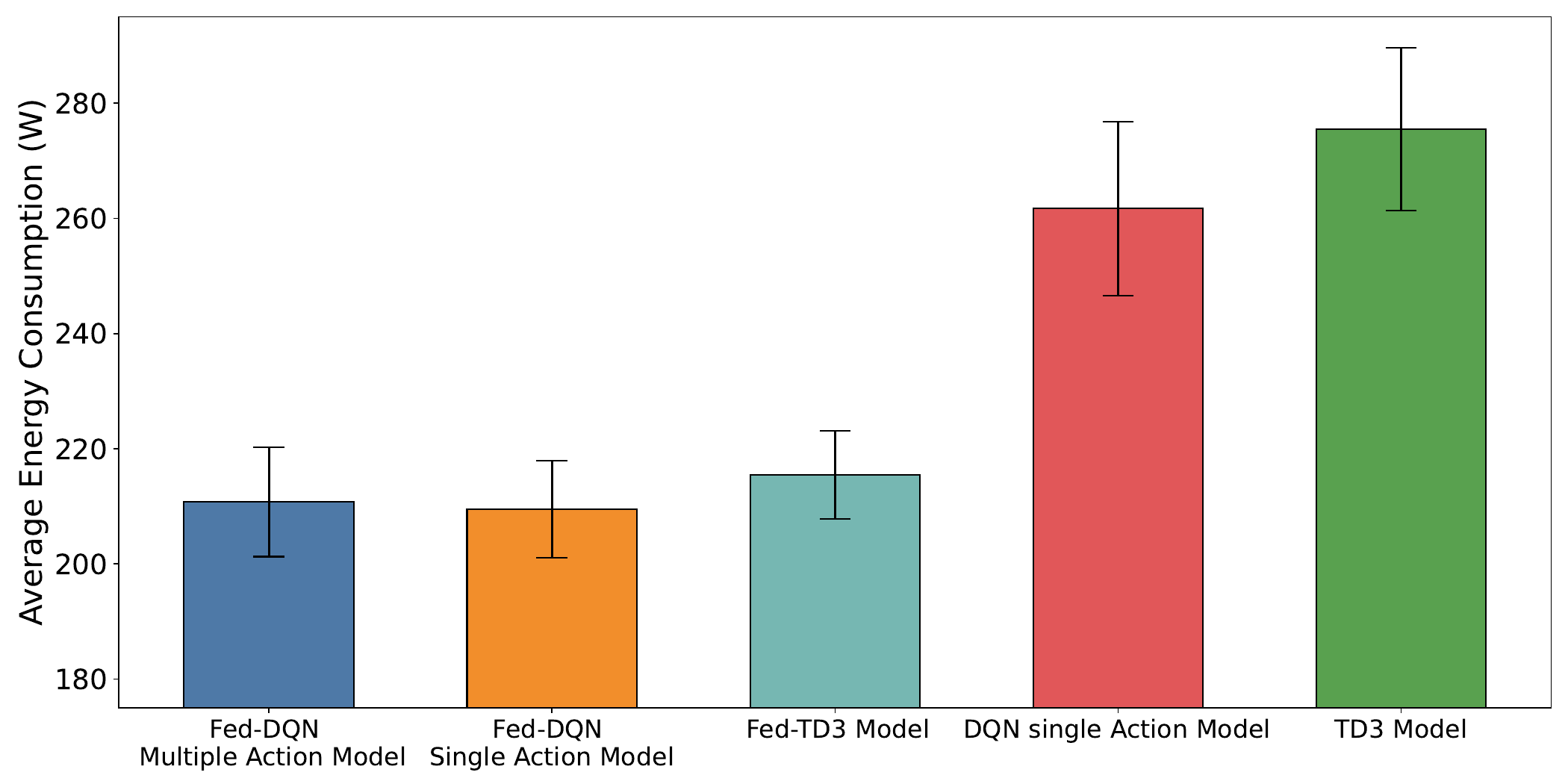}
        \label{fig:Fed500_normal1000}
    \end{subfigure}
    
    \caption{Illustrates comparison of average energy consumption (right figure) and UE satisfaction (left figure) between federated and centralized reinforcement learning approaches. The first three bars represent federated models (Fed-DQN with multiple and single actions, and Fed-TD3) applied to four independent 500\,m$\times$500\,m regions with 6 RUs and 20 UEs. The last two bars show the performance of centralized models (DQN and TD3) trained on a merged 1000\,m$\times$1000\,m region with 24 RUs and 80 UEs.}
    \label{fig:two_side_by_side}
\end{figure}

Fig. \ref{fig:two_side_by_side} compares the average energy consumption between federated and centralized reinforcement learning approaches. The federated models (e.g. Fed-DQNMA, Fed-DQNSA, and Fed-TD3) are first trained across four geographically separated 500\,m$\times$500\,m regions. After convergence, the resulting global models are independently evaluated in each of the original training regions, and the total test energy consumption is obtained by summing across all four areas.
In contrast, the centralized models (DQN and TD3) are trained on a merged 1000\,m$\times$1000\,m,
environment composed of the same four subregions, with user distributions and mobility constrained within each 500\,m$\times$500\,m area, as illustrated on the third figure of Fig. \ref{fig.radiomap}. This setting ensures that both federated and centralized models are tested on identical underlying environments, allowing a fair comparison of generalization and efficiency. The results show that federated models achieve comparable or lower overall energy consumption than their centralized counterparts, despite being trained without direct access to globally aggregated data. In particular, Fed-DQNMA and Fed-DQNSA demonstrate strong energy-saving behavior, while Fed-TD3 incurs slightly higher energy usage. This difference arises because Fed-TD3 prioritizes satisfying the QoS requirements of all users, leading to fewer RUs being turned off during testing. In contrast, the fed-DQN models are more aggressive in RU deactivation, sacrificing service quality for a greater reduction in energy consumption. Nevertheless, Fed-TD3 achieves higher average rewards, indicating a better balance between energy efficiency and user satisfaction.
These results confirm that federated reinforcement learning not only enables scalable training across geographically distributed areas, but also yields robust and energy-efficient policies that generalize well when deployed in composite environments. Note that in Fig.\ref{fig:two_side_by_side}, the TD3 model performs slightly worse than the \ac{DQNSA} in the 1000\,m$\times$1000\,m area. This is primarily due to the significantly larger action space in the centralized TD3 setting, which involves 24 RUs and poses challenges even with continuous action control. It is important to note that this comparison focuses solely on the differences between FL and non-FL approaches.

\begin{figure}[!t]
	\centering
	\includegraphics[clip, trim=0.0cm 0.cm 0.0cm 0cm, width=1\columnwidth]{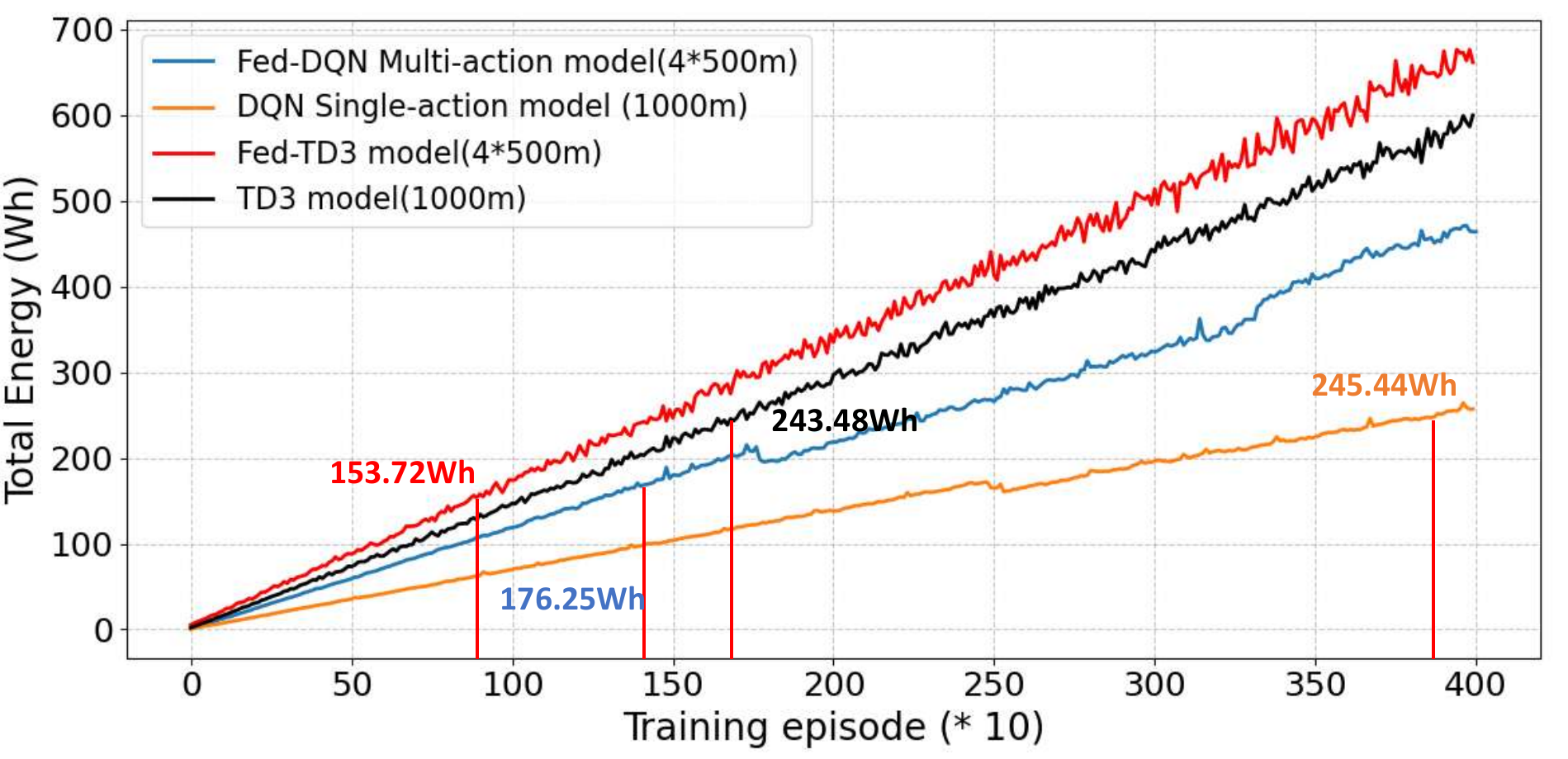}
	\caption{Illustrates training energy consumption for different DRL models. Each curve represents the cumulative energy usage over time for a specific model. The red labels indicate the total energy consumed by each model upon convergence. The comparison highlights the efficiency differences among federated and centralized approaches.}
        \label{fig:training energy consumption}
\end{figure}

Fig.~\ref{fig:training energy consumption} reports the total energy consumption required for each model to reach convergence during training. The federated models Fed-TD3 and Fed-DQNMA are compared against centralized DQN and TD3 baselines trained on the full 1000\,m$\times$1000\,m area. Each data point represents the accumulated training energy measured at the episode when the model achieves stable reward performance.
To obtain accurate measurements, we implemented a custom power monitoring module that continuously samples both CPU and GPU power usage during training. The GPU power was measured using the NVIDIA System Management Interface (nvidia-smi), while CPU utilization was monitored using the \texttt{psutil} library. Samples were collected every 100\,ms, and total energy consumption (in Wh) was computed by integrating the average power over the training duration. All models were trained exclusively on a single RTX 3060 GPU, with no other significant processes running concurrently.
Among all methods, Fed-TD3 exhibits the lowest training energy consumption at 153.72\,Wh, followed by Fed-DQNMA at 176.25\,Wh. In contrast, the centralized DRL models consume 243.48\,Wh (TD3) and 245.44\,Wh (DQN), respectively—the highest among all configurations. Notably, Fed-TD3 reduces training energy by approximately 37.4\% compared to the centralized DQN model, while Fed-DQNMA achieves a 28.2\% reduction. These results highlight the efficiency advantage of federated reinforcement learning in distributed training scenarios.


\section{Conclusions}
In this work, we explored energy-efficient RU sleep control within the O-RAN framework through both centralized and federated \ac{DRL} strategies. In the centralized setting, we evaluate DQNMA, DQNSA, and TD3, demonstrating that continuous-action control in TD3 effectively avoids the combinatorial growth of discrete methods while enabling coordinated multi-RU decisions. Building on these insights, we extend the solution to a \ac{FL} framework aligned with O-RAN’s hierarchical control, enabling scalable training across geographically distributed regions without sharing raw data. Extensive simulations across varied network scales and heterogeneous layouts show that the federated global models achieve competitive or superior performance compared to centralized training while significantly improving scalability and generalization. Compared with DQN-based baselines, Fed-TD3 achieves higher rewards, lower training energy consumption, and better adaptability to complex action spaces. These results confirm the practicality of combining centralized DRL optimization with \ac{FL} to enable flexible and energy efficiency network deployments.

\begin{acronym} 
\acro{5G}{Fifth Generation}
\acro{ACO}{Ant Colony Optimization}
\acro{ANN}{Artificial Neural Network}
\acro{BB}{Base Band}
\acro{BBU}{Base Band Unit}
\acro{BER}{Bit Error Rate}
\acro{BS}{Base Station}
\acro{BSs}{Base Stations}
\acro{BW}{bandwidth}
\acro{C-RAN}{Cloud Radio Access Networks}
\acro{CAPEX}{Capital Expenditure}
\acro{CoMP}{Coordinated Multipoint}
\acro{CR}{Cognitive Radio}
\acro{D2D}{Device-to-Device}
\acro{DAC}{Digital-to-Analog Converter}
\acro{DAS}{Distributed Antenna Systems}
\acro{DBA}{Dynamic Bandwidth Allocation}
\acro{DC}{Duty Cycle}
\acro{DL}{Deep Learning}
\acro{DSA}{Dynamic Spectrum Access}
\acro{FBMC}{Filterbank Multicarrier}
\acro{FEC}{Forward Error Correction}
\acro{FFR}{Fractional Frequency Reuse}
\acro{FSO}{Free Space Optics}
\acro{GA}{Genetic Algorithms}
\acro{HAP}{High Altitude Platform}
\acro{HL}{Higher Layer}
\acro{HARQ}{Hybrid-Automatic Repeat Request}
\acro{HCA}{Hierarchical Cluster Analysis}
\acro{HO}{Handover}
\acro{KNN}{k-nearest neighbors} 
\acro{IoT}{Internet of Things}
\acro{LAN}{Local Area Network}
\acro{LAP}{Low Altitude Platform}
\acro{LL}{Lower Layer}
\acro{LoS}{Line of Sight}
\acro{LTE}{Long Term Evolution}
\acro{LTE-A}{Long Term Evolution Advanced}
\acro{MAC}{Medium Access Control}
\acro{MAP}{Medium Altitude Platform}
\acro{MDP}{Markov Decision Process}
\acro{ML}{Machine Learning}
\acro{MME}{Mobility Management Entity}
\acro{mmWave}{millimeter Wave}
\acro{MIMO}{Multiple Input Multiple Output}
\acro{NFP}{Network Flying Platform}
\acro{NFPs}{Network Flying Platforms}
\acro{NLoS}{Non-Line of Sight}
\acro{OFDM}{Orthogonal Frequency Division Multiplexing}
\acro{O-RAN}{Open Radio Access Network}
\acro{OSA}{Opportunistic Spectrum Access}
\acro{PAM}{Pulse Amplitude Modulation}
\acro{PAPR}{Peak-to-Average Power Ratio}
\acro{PGW}{Packet Gateway}
\acro{PHY}{physical layer}
\acro{PSO}{Particle Swarm Optimization}
\acro{PU}{Primary User}
\acro{QAM}{Quadrature Amplitude Modulation}
\acro{QoE}{Quality of Experience}
\acro{QoS}{Quality of Service}
\acro{QPSK}{Quadrature Phase Shift Keying}
\acro{RF}{Radio Frequency}
\acro{RL}{Reinforcement Learning}
\acro{RMSE}{Root Mean Squared Error}
\acro{RN}{Remote Node}
\acro{RRH}{Remote Radio Head}
\acro{RRC}{Radio Resource Control}
\acro{RRU}{Remote Radio Unit}
\acro{RSS}{Received Signal Strength}
\acro{SU}{Secondary User}
\acro{SCBS}{Small Cell Base Station}
\acro{SDN}{Software Defined Network}
\acro{SNR}{Signal-to-Noise Ratio}
\acro{SON}{Self-organising Network}
\acro{SVM}{Support Vector Machine}
\acro{TDD}{Time Division Duplex}
\acro{TD-LTE}{Time Division LTE}
\acro{TDM}{Time Division Multiplexing}
\acro{TDMA}{Time Division Multiple Access}
\acro{UE}{User Equipment}
\acro{UAV}{Unmanned Aerial Vehicle}
\acro{USRP}{Universal Software Radio Platform}
\acro{DRL}{Deep Reinforcement Learning}
\acro{AI}{Artificial Intelligence}
\acro{RAN}{Radio Access Network}
\acro{RU}{Radio Unit}
\acro{CU}{Central Unit}
\acro{DU}{Distributed Unit}
\acro{NR}{New Radio}
\acro{gNBs}{Next Generation Node Bases}
\acro{CP}{Control Plane}
\acro{UP}{User Plane}
\acro{FPGAs}{Field  Programmable  Gate  Arrays}
\acro{ASICs}{Application-specific Integrated Circuits}
\acro{PHY-low}{lower level PHY layer processing}
\acro{FFT}{Fast Fourier Transform}
\acro{RRC}{Radio Resource Control}
\acro{SDAP}{Service Data Adaptation Protocol}
\acro{PDCP}{Packet Data Convergence Protocol}
\acro{RLC}{Radio Link Control}
\acro{RIC}{RAN Intelligent Controller}
\acro{RICs}{RAN Intelligent Controllers}
\acro{KPMs}{Key Performence Measurements}
\acro{RT}{Real Time}
\acro{SMO}{Service Management and Orchestration}
\acro{UE}{User Equipment}
\acro{API}{Application Programming Interface}
\acro{OSC}{O-RAN Software Community}
\acro{DRL}{Deep Reinforcement Learning}
\acro{OSP}{Online Service Provider}
\acro{NIB}{Network Information Base}
\acro{SDL}{Shared Data Layer}
\acro{SLA}{Service Level Agreement}
\acro{A1AP}{A1 Application Protocol}
\acro{HTTP}{Hypertext Transfer Protocol}
\acro{SL}{Supervised Learning}
\acro{UL}{Unsupervised Learning}
\acro{RL}{Reinforcement Learning}
\acro{DL}{Deep Learning}
\acro{FDD}{Frequency-division Duple}
\acro{TDD}{Time-division Duple}
\acro{LSTM}{Long Short-term Memory}
\acro{PCA}{Principal Component Analysis}
\acro{ICA}{Independent Component Analysis}
\acro{MDP}{Markov Decision Process}
\acro{GRL}{Generalization Representation Learning}
\acro{SRL}{Specialization Representation Learning}
\acro{SVM}{Support Vector Machine}
\acro{TDNN}{Time-delay Neural Network}
\acro{LSTM}{Long Short-term Memory}
\acro{MSE}{Mean Squared Error}
\acro{CNN}{Conventional Neural Network}
\acro{NAS}{Neural Architecture Search}
\acro{SDS}{Software Defined Security}
\acro{SON}{Self-organized Network}
\acro{KPIs}{Key Performance Indicators}
\acro{HetNet}{Heterogeneous Network}
\acro{HPN}{High Power Node}
\acro{LPNs}{Low Power Nodes}
\acro{QL}{Q-Learning}
\acro{PG}{Policy Gradient}
\acro{A2C}{Actor-Critic}
\acro{TD}{Temporal Defence}
\acro{SHAP}{SHapley Additive exPlanations}
\acro{DNNs}{Deep Neural Networks}
\acro{DNN}{Deep Neural Network}
\acro{MLP}{Multiple-layer Perceptron}
\acro{RNN}{Recurrent Neural Network}
\acro{DRL}{Deep Reinforcement Learning}
\acro{DQN}{Deep Q-Learning Network}
\acro{DDQN}{Double Deep Q-Learning Network}
\acro{GNN}{Graph Neural Network}
\acro{eMMB}{enhanced mobile broadband}
\acro{URLLC}{ultra-reliable low-latency communication}
\acro{QoS}{Quality of Service}
\acro{ILP}{Integer Linear Programming}
\acro{NF}{Network Function}
\acro{VFN}{Network Function Virtualization}
\acro{NBC}{Navie Bayes Classifier}
\acro{RAT}{Radio Access Technology}
\acro{FML}{federated meta-learning}
\acro{TS}{Traffic Steering}
\acro{CQL}{Conservative Q-Learning}
\acro{REM}{Random Ensemble Mixture}
\acro{SCA}{Successive Convex Approximation}
\acro{XAI}{eXplainable Artificial Intelligent}
\acro{D-RAN}{Distributed RAN}
\acro{C-RAN}{Cloud RAN}
\acro{v-RAN}{Virtual RAN}
\acro{RRH}{Remote Radio Head}
\acro{mMTC}{massive machine-type communication}
\acro{CDMA}{Code Division Multiple Access}
\acro{TDMA}{Time Division Multiple Access}
\acro{OFDMA}{Orthogonal Frequency-Division Multiple Access}
\acro{RRM}{Radio Resources Management}
\acro{NFV}{Network Function Virtualization}
\acro{D-RAN}{Distributed-RAN}
\acro{V-RAN}{Vritualized-RAN}
\acro{PA}{Power Amplifier}
\acro{SINR}{Signal to interference plus noise ratio}
\acro{mmWave}{millimeter Wave }
\acro{LOS}{Line of Sight}
\acro{NLOS}{Non-Line of Sight}
\acro{FSPL}{Free Space Path Loss}
\acro{EEMP}{Energy Efficiency Maximization Problem}
\acro{QFMEE}{QoS First Maximum EE}
\acro{SWES}{Switching-on/off based Energy Saving}
\acro{APC}{Area Power Consumption}
\acro{EE}{Energy Efficiency}
\acro{Near-RT}{Near-Real Time}
\acro{Non-RT}{Non-Real Time}
\acro{Open-RAN}{Open Radio Access Network}
\acro{Near-RT RIC}{Near Real Time RIC}
\acro{Non-RT RIC}{Non Real Time RIC}
\acro{TD3}{Twin Delayed Deep Deterministic Policy Gradient}
\acro{RC}{Radio Card}
\acro{UMa}{Urban Microcell path loss}
\acro{PRB}{Physical Resource Block}
\acro{API}{Application Programming Interface}
\acro{RSS}{Received Signal Strength}
\acro{DL}{downlink}
\acro{UL}{uplink}
\acro{DDPG}{Deep Deterministic Policy Gradient}
\acro{TD}{Temporal Difference}
\acro{RSRP}{Reference Signals Received Power}
\acro{FL}{Federated Learning}
\acro{UMi}{Urban Microcel}
\acro{BN}{Batch Normalization}
\acro{SBS}{Small Base Station}
\acro{FRL}{Federated Reinforcement Learning}
\acro{DQNSA}{DQN-Single Action}
\acro{DQNMA}{DQN-Multiple Action}
\acro{CSI}{Channel State Information}
\acro{AWGN}{Additive White Gaussian Noise}
\acro{RIS}{Reconfigurable Intelligent Surface}
\acro{PPO}{Proximal Policy Optimization}
\acro{UDN}{Ultra Dense Network}
\acro{ES}{Energy Saving}
\acro{MPC}{Model Predictive Control}
\end{acronym}

\bibliographystyle{IEEEtran}

\bibliography{References.bib}
\end{document}